\documentclass[aps,prx,twocolumn,footinbib,
showpacs,showkeys,longbibliography,superscriptaddress]{revtex4-2}
\usepackage{graphicx}
\usepackage{indentfirst}
\usepackage{physics}
\usepackage{braket}
\usepackage{float}
\usepackage{amsmath}
\usepackage{amssymb}
\usepackage{verbatim}
\usepackage{wasysym}
\usepackage{epstopdf}
\usepackage{CJK}
\usepackage{esint}
\usepackage{color}
\usepackage{xcolor}
\usepackage{epsfig}
\usepackage{subfigure}

\usepackage{amsfonts}
\usepackage{footmisc}
\usepackage{scrextend}
\usepackage{multirow}
\usepackage{mathtools}

\usepackage{xr-hyper}
\usepackage[hyperfootnotes=false]{hyperref}

\usepackage[english]{babel}
\usepackage{url}
\usepackage{bm}
\definecolor{darkblue}{rgb}{0,0,0.5}
\hypersetup{
colorlinks=true,
linkcolor=black,
filecolor=black,
citecolor=darkblue,
urlcolor=darkblue,
}

\newcommand{\defeq}{\vcentcolon=}

\newcommand{\vep}{\varepsilon}
\DeclareMathOperator*{\argmax}{arg\,max}
\DeclareMathOperator*{\argmin}{arg\,min}
\DeclareMathOperator*{\argX}{arg\,\textit{X}}

\newcommand\mc[1]{\mathcal{#1}}

\newcommand\bs[1]{\boldsymbol{#1}}
\newcommand\bff[1]{\text{\textbf{#1}}}

\newtheorem{defin}{Definition}

\usepackage{algorithm}%thttp://ctan.org/pkg/algorithms
\usepackage[noend]{algpseudocode}% http://ctan.org/pkg/algorithmicx

\makeatletter
\def\algbackskip{\hskip-\ALG@thistlm}
\makeatother

\newcommand{\SP}[1]{{{\textcolor{black}{#1}}}}
\newcommand{\CH}[1]{{{\textcolor{black}{#1}}}}

\begin{document}

\title{Practical Routing and Criticality in\\ Large-Scale Quantum Communication Networks}
\author{Cillian Harney}
\affiliation{Department of Computer Science, University of York, York YO10 5GH, UK}
\affiliation{nodeQ, 71-75 Shelton Street, Covent Garden, London WC2H 9JQ, UK}
\author{Stefano Pirandola}
\thanks{Corresponding author}
\email{stefano.pirandola@york.ac.uk}
\affiliation{Department of Computer Science, University of York, York YO10 5GH, UK}
\begin{abstract}
The efficacy of a communication network hinges upon both its physical architecture and the protocols that are employed within it. In the context of quantum communications, there exists a fundamental rate-loss tradeoff for point-to-point quantum channels such that the rate for distributing entanglement, secret keys, or quantum states decays exponentially with respect to transmission distance. Quantum networks are the solution to overcome point-to-point limitations, but they simultaneously invite a challenging open question: How should quantum networks be designed to effectively and efficiently guarantee high rates? Now that performance and physical topology are inexorably linked, this question is not easy, but the answer is essential for a future quantum internet to be successful. 
In this work, we offer crucial insight into this open question for complex optical-fiber quantum networks. Using realistic descriptions of quantum networks via random network models and practical end-to-end routing protocols, we reveal critical phenomena associated with large-scale quantum networks. Our work reveals the weaknesses of applying single-path routing protocols within quantum networks, observing an inability to achieve reliable rates over long distances. Adapting novel algorithms for multi-path routing, we employ an efficient and practical multi-path routing algorithm capable of boosting performance while minimizing costly quantum resources.
\end{abstract}
\maketitle

\section{Introduction}
The deployment of a world-wide quantum network will enable provably-secure communication and cryptography \cite{KimbleQI, UniteQInt, RazaviQNet, AdvCrypt,Micius2021,AdvSQC}, and facilitate distributed quantum computing  \cite{DQC_Ekert,QN_teles,QN_clocks} on a global scale. A quantum internet will provide new means of information processing, leading to enhancements in a number of technological domains \cite{Mike_Ike, WatrousTxt, Holevo19}.

However, in many ways a quantum internet will not be like its classical counterpart \cite{SlepianNets, Satorras04, CoverThomas, TanenbaumNets, GamalNets}. Unlike classical information which is robust and easily copied, quantum information is fragile and the laws of quantum mechanics prohibit cloning. As a result, the rate at which one can perform entanglement distribution, quantum key distribution (QKD) or quantum state transfer is fundamentally limited by the medium through which it travels. The optimal transmission rate through optical fiber (the primary conduit for communications) is known exactly as the Pirandola-Laurenza-Ottaviani-Banchi (PLOB) bound~\cite{PLOB} which decays exponentially with respect to transmission distance~\cite{PirPatron09}. Crucially, the infrastructure necessary for reliable quantum communications are not immediately compatible with current classical constructions, \SP{motivating numerous contributions in the literature~\cite{PirRepAss,End2End,pant2019routing,meng2021concurrence,meng2025path,caleffi2017optimal,shi2020concurrent,fischer2021distributing,kolar2022adaptive,huang2025peer}.}

The fundamental limits of \textit{end-to-end} communication in quantum networks are also well understood \cite{End2End}. Optimal performance in an end-to-end setting is achieved by a network with capacity achieving links operating a flooding protocol; a protocol in which {every} channel in the network is used to facilitate communication between two end-users. These tools have been explored and expanded in a number of works, including the development of analytical tools for large-scale networks \cite{OPGQN, FSEnd2End, IRQNs} and experimental demonstrations in QKD networks \cite{ScalAuth}. Zhuang \textit{et al}.~carried out invaluable assessments of random, optical-fiber quantum networks \cite{QuntaoRandQNets, ZhangQInt}. Random network models are capable of capturing complex behaviours that arise within real world architectures \cite{RandomNets,HandbookRandNet}. Hence, understanding how quantum communications can be performed effectively across such models provides insight into desirable properties for the quantum internet. 

Nonetheless, glaring gaps exist in our understanding of realistic performance and resource demands. 
The most expensive resource in a quantum internet will be the number of quantum repeaters needed in a physical network area to guarantee effective communications, i.e.~the network nodal density. While classical repeaters are cheap and easy to deploy, quantum repeaters are costly and should not be wasted as non-user nodes. Recent studies have been able to identify important nodal density conditions for reliable end-to-end performance on quantum networks \cite{BritoRandQNets,QuntaoRandQNets,OPGQN}. Of these investigations, Ref.~\cite{BritoRandQNets} principally focuses on network connectivity properties and a specific single-photon transmission protocol through pure-loss channels, while Refs.~\cite{QuntaoRandQNets,OPGQN} evaluate network capacities with a protocol agnostic approach, using pure-loss channels and optimal flooding protocols. 

There is room for progress in each of these works. Network connectivity investigations are useful, but are not sufficient to benchmark end-to-end performance. Beyond pure-loss models, optical fiber is most accurately described using thermal-loss channels that account for environmental and experimental thermal noise. Finally, the employment of flooding protocols on a large-scale is highly impractical. It is not realistic to assume the use of \textit{every} edge in a potentially global network to facilitate communication between a single end-user pair. Some of these assumptions are effective and completely suitable for network capacity assessments, they are over-optimistic in the context of revealing stricter, practical insight for quantum networking. 

In this work, we tighten the assessment of critical resource requirements in realistic quantum networks with two crucial improvements. Firstly, we consider random network architectures composed of realistic link-layers; using point-to-point channels which account for thermal noise, experimental imperfections and practical point-to-point protocols. 
This immediately improves our characterisation of realistic quantum networks, and reveals the impact of point-to-point limitations on network connectivity models. 

Secondly, we explore the efficacy of practical end-to-end routing in quantum networks to identify critical resource requirements of quantum networks. Moving away from impractical flooding schemes, we study the feasibility of single-path routing and resource efficient multi-path routing, adopting recent results in the efficient generation of end-to-end multi-paths \cite{1SMDP2}. Our results suggest that single-path routing is an inefficient strategy for reliable rates on large-scale quantum networks. Instead, there exist efficient multi-path routing protocols which can efficiently achieve close to optimal rates without consuming the entire network. Furthermore, different network architectures observe a suitability to quantum multi-path routing which others do not. All of these results identify important design criteria for future, large-scale quantum networks while motivating the further development of practical multi-path protocols.

\section{Quantum Networks\label{sec:Prelims}}

\subsection{Preliminaries}
A quantum communication network can be described as a collection of nodes $P = \{ \bs{x}_i \}$ which are interconnected by a collection of edges $E =\{(\bs{x}_i, \bs{x}_j)\}_{i,j}$, resulting in a finite, undirected graph $\mc{N} = (P,E)$. 
A network node $\bs{x} \in P$ represents a station which contains a register of quantum systems and the capability to transmit these systems to other nodes in the network. This station may act as a repeater node, or a potential user of the network. 
A pair of nodes $\bs{x}$ and $\bs{y}$ are connected by an undirected edge $(\bs{x},\bs{y})\in E$ if there exists a quantum channel $\mc{E}_{\bs{xy}}$ between them; permitting the exchange of quantum systems between them. 

A quantum network can be used to facilitate many different communication configurations which fall into two main categories: End-to-end and multi-end communications. In an end-to-end network protocol, two remote users $\bs{a},\bs{b} \in P$ aim to establish quantum communication between one another, and all other nodes in the network can be used to enable this goal. Multi-end communication configurations form a much broader class with tasks ranging from multiple-unicast (many pairs of end-to-end protocols simultaneously executed on the same network) to multi-casting (senders communicating unique messages with many receivers) and broadcasting (senders communicating identical messages with many receivers). 

In this work, we focus on end-to-end strategies. The ability of a quantum network to achieve strong end-to-end rates serves as a primitive for all other communication configurations. If strong end-to-end rates cannot be achieved, this promises difficulties for more complex multi-end scenarios in which network competition will only serve to reduce performance. 

% \ifx
% It is important to make a distinction between \textit{physical} and \textit{logical} information flow. The logical direction of a quantum communication channel refers to the direction in which entanglement, secret-keys, or quantum states are exchanged from node $\bs{x}$ to node $\bs{y}$ (or vice versa). Physical flow refers to the direction of quantum system exchange, i.e.~if quantum systems are physically sent from a node $\bs{x}$ to the node $\bs{y}$. In general, a quantum channel $\mc{E}_{\bs{x}\rightarrow\bs{y}}$ which physically transmits quantum systems from node $\bs{x}\rightarrow\bs{y}$ may be different from its reverse channel $\mc{E}_{\bs{y}\rightarrow\bs{x}}$. But thanks to quantum teleportation, either direction of logical information flow can be achieved by using any physical channel direction; hence, it is always possible to choose the \textit{best} physically directed channel for logical quantum communication (e.g.~that which has the largest capacity). \footnote{For the purposes of QKD, the resolution of physical and logical flow is even simpler; the end-goal of this protocol is symmetric, i.e.~both Alice and Bob will result in a secret-key, hence the original physical communication direction is irrelevant.}.
% \fi

\subsection{Network Link Layers}

For an arbitrary quantum network $\mc{N} = (P,E)$ we can define a rate distribution $\mc{K}$ as the collection of point-to-point rates $\mc{K} \defeq \{K_{\bs{xy}} \}_{(\bs{x},\bs{y})\in E}$, where $K_{\bs{xy}}$ denotes the rate at which a network edge is able to operate. The rate distribution of any network depends on the end-to-end task at hand (key distribution, entanglement distribution, quantum state transfer), the channel quality and the point-to-point protocols being used. Here, we focus on realistic description of the link layer via bosonic thermal-loss channels and consider two distinct descriptions of their rates: (\textit{i}) Each link operates at its two-way assisted quantum and private capacities, delivering point-to-point protocol-agnostic insight to the limits of the network, and: (\textit{ii}) The rate of each link is described by an asymptotic secret key-rate according to practical QKD protocols, providing insight into the properties of future QKD networks. 

We begin by considering quantum channel capacities. Bosonic quantum communications performed over optical-fiber is best described using a Gaussian thermal-loss channel $\mc{E}_{\eta,\bar{n}}$ of transmissivity $\eta \in (0,1)$ and output thermal noise of $\bar{n}$ photons. This channel can be described by the action of a beam-splitter of transmissivity $\eta$ which mixes the input mode with an environmental thermal mode with mean photon number ${\bar{n}_{\text{env}} \defeq \bar{n}/(1-\eta)}$~\cite{endnote_excess}.
 This transforms the input quadratures of a single-mode input Gaussian state ${\hat{\bff{x}} = (\hat{q},\hat{p})^T}$ according to $\hat{\bff{x}} \rightarrow \sqrt{\eta}\hat{\bff{x}} + \sqrt{1-\eta}\hat{\bff{x}}_{\text{env}}$, where $\hat{\bff{x}}_{\text{env}}$ is the quadrature operator of the thermal environmental mode. When $\bar{n} = 0$, the environmental mode is in the vacuum state, and this becomes a pure-loss channel.

Unfortunately, the exact capacity of bosonic thermal-loss channels is not known but can instead be tightly bounded. 
Defining the channel transmissivity ${\eta_{\text{ch}} \defeq 10^{-\gamma d}}$ where $\gamma \approx 0.02$ (corresponding to $0.2$ dB/km) is the fiber loss rate and $d$ is the channel length (km), we can write the bounds
${\mc{T}_{\eta,\bar{n}}^{\>{l}} \leq \mc{C}(\mc{E}_{\eta,\bar{n}}) \leq \mc{T}_{\eta,\bar{n}}^{\>u}}$,
where we define the bounding functions,
\begin{align}
\mc{T}_{\eta,\bar{n}}^{\>{l}} &\defeq -\log_2(1-\eta) - h\left( {\bar{n}_{\text{env}}} \right),\\
\mc{T}_{\eta,\bar{n}}^{\>{u}} &\defeq \mc{T}_{\eta,\bar{n}}^{\>{l}}  - {\bar{n}_{\text{env}}} \log_2(\eta),
\end{align}
and $h(x) \defeq (x+1)\log_2(x+1) - x\log_2(x)$ is an entropic function. 

Hence, the optimal rate of any end-to-end protocol on a bosonic thermal-loss network can always be bounded by modelling its rate distributions using the single-edge capacity bounds. That is, given a network $\mc{N} = (P,E)$, we may investigate the performance of an end-to-end protocol on this network subject to the rate distributions,
\begin{align}
\mc{K}_{l} &= \{ \mc{T}_{\eta_{\bs{xy}}, \bar{n}_{\bs{xy}}}^{\>l}\}_{(\bs{x},\bs{y})\in E}, \label{eq:RD_l} \\
\mc{K}_{u} &= \{ \mc{T}_{\eta_{\bs{xy}}, \bar{n}_{\bs{xy}}}^{\>u}\}_{(\bs{x},\bs{y})\in E}. \label{eq:RD_u}
\end{align}
Using this knowledge, we can bound the optimal performance of end-to-end protocols on bosonic thermal-loss networks. A similar approach can of course be applied to networks composed of other channels with unknown channel capacities.

Unfortunately, practical, near-term performance levels for quantum communications remain some way off of optimal rates described by channel capacities. It is therefore beneficial to investigate the properties of quantum networks subject to realistic protocols and technologies that are achievable today. In this regard, QKD stand as the most deployable quantum communication task to date; yet, realistic resource assessments of QKD networks remain largely unknown. 
Hence, we also initiate an assessment of large-scale trusted-node QKD networks. By treating each network node as a trusted user station, we are able to investigate large-scale network constructions in which the single-edge rates are captured by well known point-to-point secret-key rates. In this way, we are able to exploit practical key-rates associated with QKD protocols and assess the capabilities/resource requirements of QKD networks. 
To see the precise key-rate expressions, we point the reader to Appendix \ref{sec:KeyRates}.

\subsection{End-to-End Routing and Rates \label{sec:E2ERates}}

An end-to-end protocol is characterised by its routing strategy, the protocol which dictates the path (or paths) that logical quantum communication~\cite{Note1} follows throughout the network in order to communicate between end-users. An end-to-end route between a pair of remote end-users $\bs{a}$, $\bs{b}$ can be defined as a collection of network edges which connect one end-user to the other. We can define a single-path route as a set
\begin{equation}
\omega \defeq \{(\bs{a},\bs{x}_1), (\bs{x}_1,\bs{x}_2),\ldots,(\bs{x}_{N},\bs{b})\},
\end{equation}
and define a route as free of cycles (without loss of generality). Any route $\omega$ refers to an associated collection of quantum channels
$
\{ \mc{E}_{\bs{xy}} \}_{(\bs{x},\bs{y}) \in \omega} = \{ \mc{E}_{\bs{a}\bs{x}_1},  \mc{E}_{\bs{x}_1\bs{x}_2}, \ldots, \mc{E}_{\bs{x}_N\bs{b}}\},
$
through which quantum systems must be exchanged to establish end-to-end quantum communications. In quantum networks, a routing strategy must be chosen to dictate how network interactions are followed. There exist two main classes of end-to-end routing strategy: \textit{Single-path} and \textit{multi-path} routing.

% \footnote{It is important to make a distinction between \textit{physical} and \textit{logical} information flow. The logical direction of a quantum communication channel refers to the direction in which entanglement, secret-keys, or quantum states are exchanged from node $\bs{x}$ to node $\bs{y}$ (or vice versa). Physical flow refers to the direction of quantum system exchange, i.e.~if quantum systems are physically sent from a node $\bs{x}$ to the node $\bs{y}$. In general, a quantum channel $\mc{E}_{\bs{x}\rightarrow\bs{y}}$ which physically transmits quantum systems from node $\bs{x}\rightarrow\bs{y}$ may be different from its reverse channel $\mc{E}_{\bs{y}\rightarrow\bs{x}}$. But thanks to quantum teleportation, either direction of logical information flow can be achieved by using any physical channel direction; hence, it is always possible to choose the \textit{best} physically directed channel for logical quantum communication (e.g.~that which has the largest capacity).} 

Single-path routing describes a network protocol, $\mc{P}_{\text{sp}}$, in which quantum systems are exchanged from node-to-node throughout the network in a sequential manner. This forges a unique path of interactions through the network, and continues until quantum communication has been established between the end-users. This is the standard strategy used for classical networking and is extremely effective thanks to the robustness of classical information. 

The vulnerability of quantum networks to decoherence means that it is extremely valuable to explore more general multi-path protocols, $\mc{P}_{\text{mp}}$ in order to enhance performance. Network nodes may exchange multiple quantum systems with many receiver nodes, repeating until communication is established between the end-users. End-users may explore a variety of end-to-end routes simultaneously in a way which enhances their performance.

Let us give a generalised description of an end-to-end protocol. Consider a quantum network $\mc{N} = (P,E)$, a routing protocol $\mc{P}$ and an end-user pair Alice $\bs{a}$ and Bob $\bs{b}$ which we collect into a single end-user quantity for ease of notation, $\bs{i} \defeq \{ \bs{a},\bs{b} \}$ ($\bs{a},\bs{b} \in P$). Alice prepares a generally multipartite quantum state in her register and performs a point-to-multipoint transmission, exchanging quantum systems with all the nodes in her neighbourhood $\bs{a} \rightarrow n_{\bs{a}} \defeq \{\bs{x}\in P~|~(\bs{a},\bs{x})\in E\}$,
where $ n_{\bs{x}}$ denotes the neighbourhood of a node $\bs{x}$. We assume that the use of any edge is probabilistic, such that any node $\bs{x}$ only exchanges systems with a node $\bs{y}$ according to a \textit{forwarding probability}, $q_{\bs{xy}} \in [0,1]$. After this point-to-multipoint transmission, each node $\bs{x} \in n_{\bs{a}}$ will do the same thing with each of their neighbours and perform system exchanges along any \text{available edge} which has not yet been used (according to their own forwarding probability). This process of multipoint exchanges will continue throughout the network until eventually a final multipoint-point exchange will occur with Bob's node and his neighbourhood $n_{\bs{b}} \rightarrow \bs{b}$. At this point, communication has been established between the end-users. 

Over the course of many transmissions, every edge ${(\bs{x},\bs{y})\in E}$ in the network will have been utilised $q_{\bs{xy}}$-times on average. In this way, we have described a generalised version of the standard \textit{flooding protocol} $\mc{P}_{\text{fl}}$ \cite{End2End}. Instead of using every edge once in order to establish end-to-end communication, i.e.~$q_{\bs{xy}} =  1$ for all ${(\bs{x},\bs{y})\in E}$, each edge is used $q_{\bs{xy}} \leq 1$ times on average. 

In large networks, the use of standard flooding is not practical, since this requires the use of potentially huge number of edges for a single-pair of uses.
The distribution of forwarding probabilities $\{q_{\bs{xy}} \}_{(\bs{x},\bs{y}) \in E}$ is completely defined by the chosen routing protocol.
It can be readily modified to exclusively perform end-to-end communication along a selection of $M \geq 1$ paths which we collect into the route set, $\Omega_{\mc{P}} \defeq \{\omega_1, \omega_2, \ldots, \omega_M\}$. It is useful to denote an analogous routing edge set $E_{\mc{P}}$ which unpacks the route set into all its unique edges, 
\begin{equation}
E_{\mc{P}} \defeq  \omega_1 \cup \ldots \cup \omega_M = \bigcup_{i=1}^M \omega_i.
\end{equation}
More precisely, a protocol $\mc{P}$ which deterministically uses a finite set of routes $\Omega_{\mc{P}}$ generates a forwarding probability distribution of the form
\begin{equation}
q_{\bs{xy}} = \begin{cases}
1 & \text{ iff }  (\bs{x},\bs{y}) \in E_{\mc{P}}, \\
0 & \text{ otherwise}, \label{eq:Path_Flooding}
\end{cases}
\end{equation}
for all $(\bs{x},\bs{y})\in E$. It is clear that if we reduce this set of potential paths to only one route $\Omega_{\mc{P}} = \{\omega_1\}$ then we reclaim a single-path protocol, $\mc{P}_{\text{sp}}$. Otherwise, we engage in a multi-path protocol, $\mc{P}_{\text{mp}}$.

The end-to-end rate associated with this general form of protocol can be computed by solving the max-flow min-cut theorem \cite{End2End}. 
That is, an end-to-end rate is found by determining the set of edges in the network which simultaneously disconnect the end-user pair and minimize the sum of all their single-edge rates. We call this partitioning a \textit{network cut} $C$ which generates the set of edges which creates the partition, or cut-set $\tilde{C}$ (see Appendix \ref{sec:CutSet}). Given a single-edge rate $K_{\bs{xy}}$ and its forwarding probability $q_{\bs{xy}}$, its effective rate is given by $q_{\bs{xy}} K_{\bs{xy}}$.
Hence, the end-to-end rate between $\bs{i}$ on the network $\mc{N}$ using $\mc{P}$ can be computed by
\begin{equation}
{K} (\bs{i}, \mc{N} | \mc{P}) \defeq \min_{C} \sum_{(\bs{x},\bs{y})\in \tilde{C}}  q_{\bs{xy}} {K}_{\bs{xy}} \label{eq:GenFlood}.
\end{equation}

\section{Random Quantum Networks \label{sec:RandNets}}

\subsection{Link Quality and Edge Pruning}
The study of random network models is a rich and wide-spanning field, within which many key tools have been developed. Of course, random network models cannot capture \textit{all} of the features of a future quantum internet; these features will emerge as the technology is developed and deployed. Nonetheless, relevant random networks are able to capture and predict important behaviours of realistic complex networks. In this work, we consider the classes of Waxman and scale-free networks with two important considerations: Realistic link-layer descriptions and practical end-to-end protocols. 

To address the weak rates achieved by quantum links over long distances, we impose a modification to the standard random network models. Network edges which possess a point-to-point rate below some threshold value $\vep$ are \textit{pruned} and removed from the network edge set after graph generation. Under this modification networks can only be considered completely connected if they are \textit{competently connected} by edges with $K_{\bs{xy}} \geq \vep$. Given that desirable clock rates $C$ for quantum communications are on the order of $\text{GHz}$, we set $\vep \sim 10^{-12}$ so that valid network edges must guarantee at least $CK_{\bs{xy}} \gtrsim 1$ mbit per second~\cite{Note2}.
%\footnote{Note that is already a lenient condition, as clock rates in many quantum communication protocols are thus far restricted to the MHz range. Furthermore, the collapse of the point-to-point thermal loss rate functions mean that the minimum length of pruned edges are very similar for threshold values as large as $\vep \sim 10^{-6}$, so this choice is quite flexible.}.
The notion of pruning goes towards preserving network resources that are otherwise wasted, and offers a more accurate representation of network connectivity. Importantly, it affects our random network models in unique ways, reflecting the potential impact of poor link quality in a realistic quantum internet. 

\begin{figure}
\includegraphics[width=0.95\linewidth]{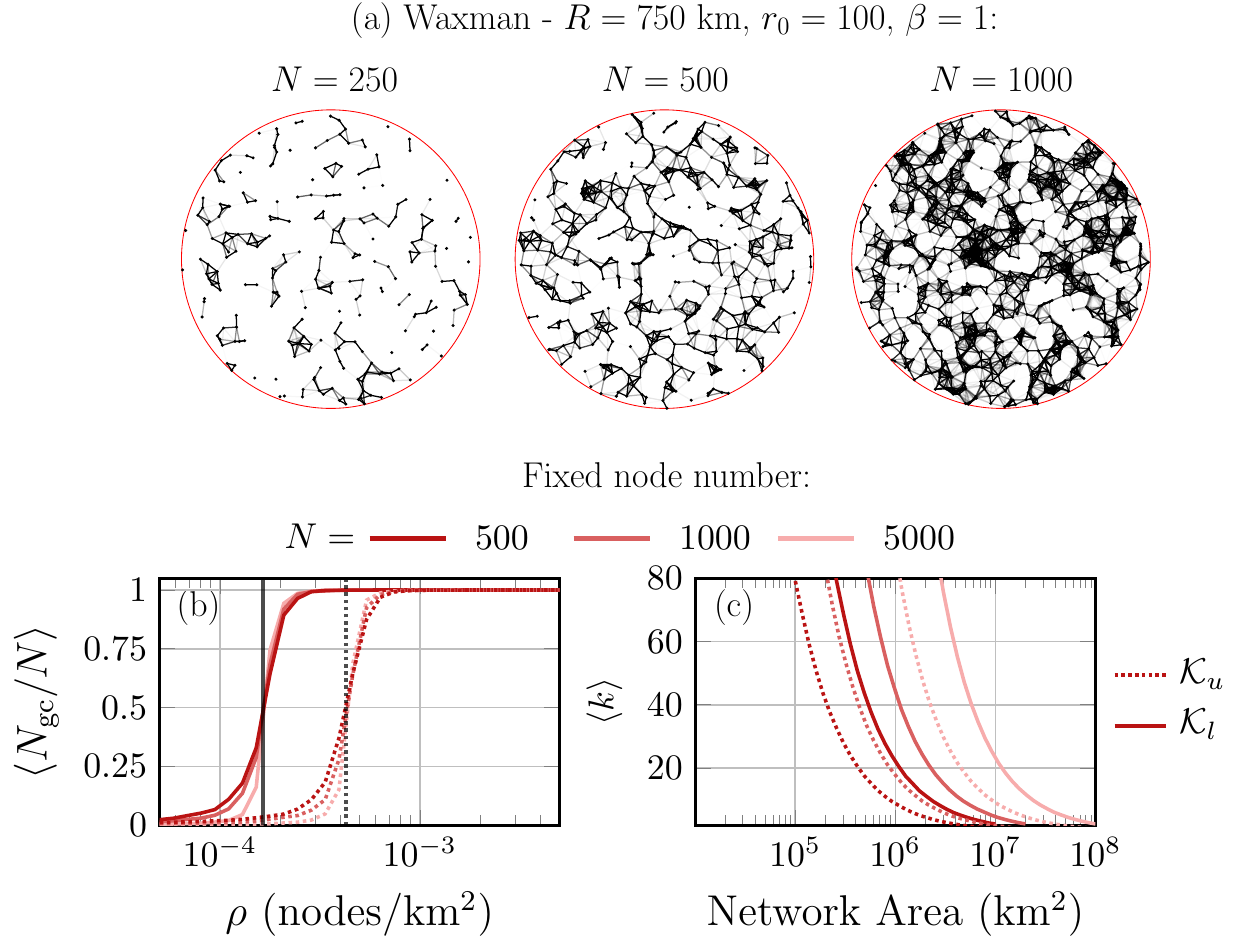}
\caption{Connectivity properties of bosonic thermal-loss Waxman networks. Panel (a) displays random networks generated under the parameters listed above and using single-edge capacity upper-bounds. The colour intensity of each edge is proportional to its capacity.
Panel (b) plots the average main component fraction, in which the critical densities necessary for a connectivity phase transition are identified by the black lines. Panel (c) shows the average degree of networks with a fixed number of nodes (given in legend) with respect to variable network density and area respectively. The legend identifies the fixed node number and indicates whether the network uses upper or lower bounds on the rate distributions. }
\label{fig:WaxConn}
\end{figure}

\subsection{Waxman Networks}
Networks from the Waxman class $\mc{N} \in \textsf{W}$ are constructed as follows: A number of $N$ nodes are generated in a region of area $A$, and any pair of nodes $\bs{x}, \bs{y}$ are connected with a probability that decays exponentially with respect to their point-to-point separation, $r_{\bs{xy}}$. More precisely, a channel $\mc{E}_{\bs{xy}}$ between the nodes $\bs{x},\bs{y}$ is created with probability 
\begin{equation}
p_{\bs{xy}}^{\textsf{W}} \defeq \beta e^{-\frac{r_{\bs{xy}}}{r_0}}.  \label{eq:WaxProb}
\end{equation}
The parameters $\beta, r_0$ are characteristic of the model and influence generation process; $\beta \in (0,1]$ defines a maximum probability of connection, while $r_0 \in (0,\infty)$ dictates the speed at which the connection probability exponentially decays.
Erd\H{o}s-R\'{e}nyi networks are a specification of the Waxman model such that $r_0 \rightarrow \infty$, i.e.~there is no decay with respect to channel length but the probability of connection is simply $p_{\bs{xy}}^{\text{ER}}  \defeq \beta \leq 1$. 

Examples of $N$ node networks contained in circular areas of radius $R$ can be seen in Fig.~\ref{fig:WaxConn}(a). In order to analyse these networks, we define $N_{\text{gc}}$ as number of nodes in the giant component of the network (largest subset of nodes between which all nodes possess paths to one another). Furthermore, we define the degree of a node $k_{\bs{x}}$ as the number of nodes to which it is connected. 
When $N$ is very small (or $R$ is very large) the network is sparse and poorly connected; the average distance between nodes is large so that links will connect with very low probability under Eq.~(\ref{eq:WaxProb}). As a result, the network will be clustered and incompletely connected~\cite{Note3}.
%\footnote{The number of nodes in the giant component is less than the number of network nodes $N$}. 
This can be seen in Fig.~\ref{fig:WaxConn}(a) for $N = 250$ nodes. Yet, as $N$ is increased (or $R$ is decreased) then the average nodal separation will shrink, increasing the likelihood of connections. Eventually, the nodal density becomes large enough to promise that the number of nodes in the giant component $N_{\text{gc}} = N$ and there exists end-to-end paths between all nodes. 

This behaviour in Waxman networks is well known, and gives rise to a percolation phase transition. As seen in Fig.~\ref{fig:WaxConn}(b) There exists a critical nodal density $\rho_{\text{G}}^\star$ at which the model abruptly transits from poorly connected ($N_{\text{gc}} < N/2$) to well connected ($N_{\text{gc}} \geq N/2$). The phase transition is illustrated in Fig.~\ref{fig:WaxConn}(b) in the context of bosonic thermal-loss networks, where we plot the average fraction of the network contained in the giant component, $\langle N_{\text{gc}}/N\rangle$. One can similarly see in Fig.~\ref{fig:WaxConn}(c) how the average nodal degree $\langle k \rangle$ undergoes a collapse as the network area is expanded, leading to the connectivity transition.
It can be seen that the critical density is bounded between $1.6 \times 10^{-4} \lesssim \rho_{\text{G}}^\star \lesssim 4.3 \times 10^{-4}$ nodes per km${}^2$. This is nearly two orders of magnitude larger than that predicted by bosonic pure-loss networks, for which $\rho_{\text{G}}^{\star} \sim 7\times 10^{-6}$ nodes per km${}^2$ \cite{BritoRandQNets,QuntaoRandQNets}. This emphasises thermal decoherence as a vital consideration since the impact of environmental noise alone can significantly degrade connectivity. 

\subsection{Scale-free Networks \label{sec:SFNets}}
\begin{figure}
\includegraphics[width=\linewidth]{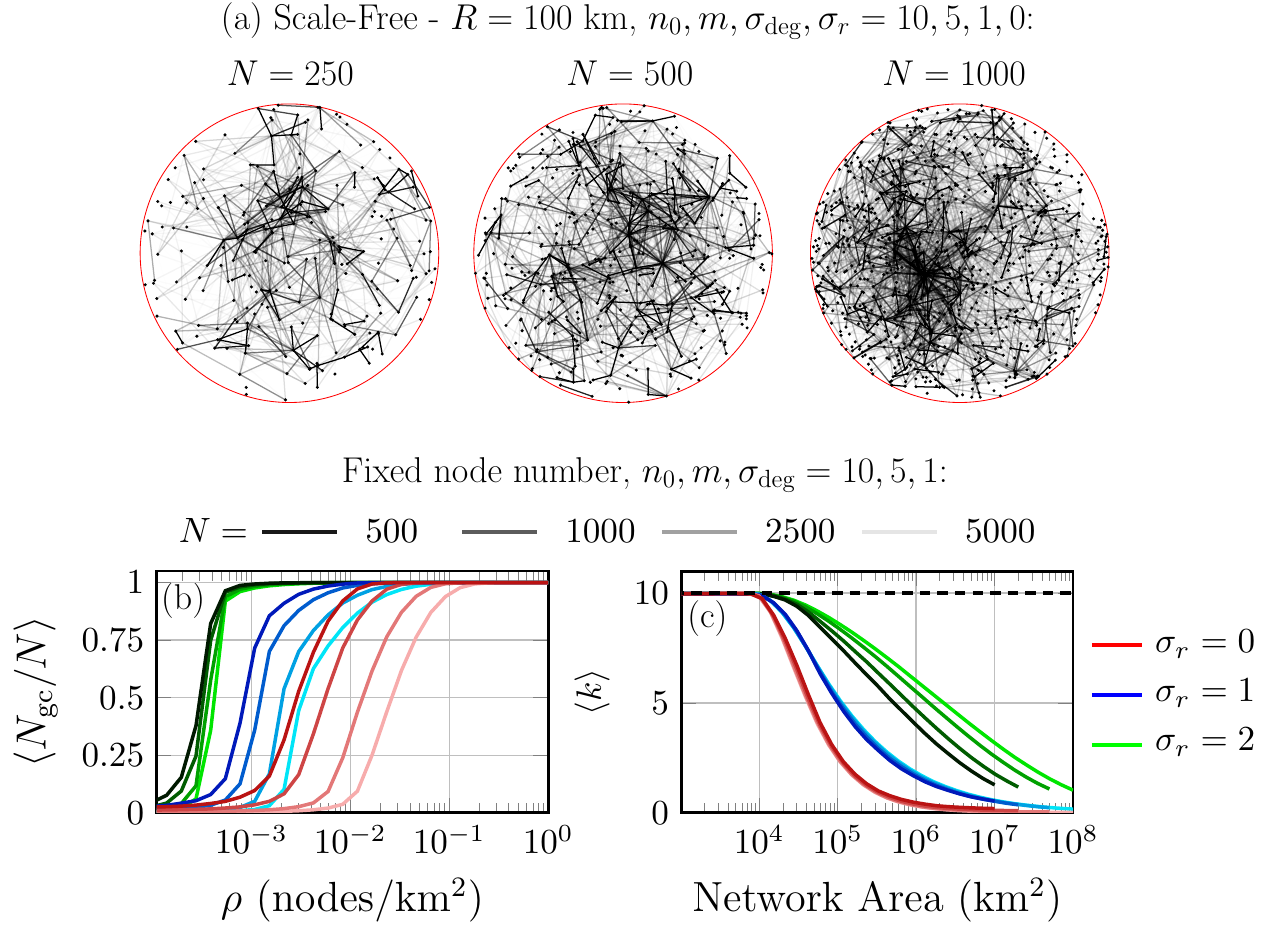}
\caption{Connectivity properties of bosonic thermal-loss scale-free networks. Panel (a) displays random networks generated under the parameters listed above and using single-edge capacity upper-bounds. The opacity of each edge is proportional to its capacity. Panel (b) plots the average giant component fraction and (c) the average degree of networks with a fixed number of nodes with respect to variable network density and area respectively. The legend identifies the fixed node number and $\sigma$ exponent.}
\label{fig:SFConn}
\end{figure}
Another important random architecture is that of scale-free networks $\textsf{S}$. A network is scale-free if their degree distribution (probability distribution of nodes in $\textsf{S}$ having degree $k$) follows a power law, i.e.~$p_k \propto k^{-\gamma}$ where $\gamma$ is some real number characterizing the distribution. Many real world networks (such as the classical internet) are thought to exhibit scale-free properties, however it is rare that a network is precisely scale-free \cite{SFNareRare}. In this work, we study scale-free networks generated dynamically using Yook's model \cite{Yook,SF} in which new nodes $\bs{y}$ are iteratively added to an initially small, $n_0$ node connected network. Each new node $\bs{y}$ is attached to a collection of $m$ existing nodes $\{\bs{x}_i\}_{i=1}^m$ with probability 
\begin{equation}
p_{\bs{xy}}^{\textsf{S}} \propto {k_{\bs{x}}^{\sigma_{\text{deg}}}}/{r_{\bs{xy}}^{\sigma_{{r}}}}, \label{eq:ProbSF}
\end{equation} 
where $\sigma_{\text{deg}}, \sigma_r \in \mathbb{R}_0^+$ are model parameters which controls the influence degree and link length have on connection probability.

Some example networks are illustrated in Fig.~\ref{fig:SFConn}(a). In general, the connectivity properties of scale-free networks are very different to those of Waxman networks. Fig.~\ref{fig:SFConn}(b) shows the behaviour of the giant component with respect to fixed node number and nodal density. When there is no connection dependence on link length ($\sigma_r=0$) there is no critical density. The giant component eventually transits to $\langle N_{\text{gc}}/N \rangle = 1$, but less abruptly, and at a higher density than that of Waxman networks. Increasing $\sigma_r$, one can see that the network becomes fully connected more quickly and a critical density begins to emerge as we recover the transition shape from Fig.~\ref{fig:WaxConn}(b). 

There are also notable differences in the context of average nodal degree, $\langle k \rangle$. The average degree of Waxman networks scales exponentially with respect to nodal density. In the scale-free setting, the relationship between connection probability and nodal degree gives rise to nodal hubs (i.e.~nodes to which most others are connected) but also a significant level of sparsity, so that most nodes have a very low degree. This is due to the mechanism of preferential attachment, as nodes which have high degrees are more likely to gain more connections. As seen in Fig.~\ref{fig:SFConn}(c), at high densities, scale-free networks saturates to a maximum value $\langle k \rangle \rightarrow 2m$. 

However, at lower network densities this is not the case and the average degree undergoes a collapse. This can be understood as follows: When new nodes are added, they attempt to connect with $m$ other existing network nodes. At low densities, nodes are typically too far away to forge competent links. When a new node is added, the $\sigma_r = 0$ model has no preference in choosing $m$ nodes within a quality connection range and thus new nodes fail to connect reliably and the average degree collapses. For $\sigma_r = 1$, there exists some preference for choosing nearby nodes which is capable of slowing the $\langle k\rangle$ decay. Forcing a stronger link length dependence at $\sigma_r = 2$ slows this rate even further and raises the degeneracy with respect to node number.

Clearly, setting the parameter $\sigma_r = 2$ breaks the ``scale-freedom" of this model, since the degree behaviour is no longer independent from $N$. However, this dependence is an inevitable consequence of utilizing quantum communications. As with the classical internet, it is likely that scale-free properties will emerge in quantum networks. However, network engineers will not turn a blind eye to link length and channel quality, making the $\sigma_r = 0$ model unrealistic, rendering larger values of $\sigma_r$ meaningful and interesting. Throughout this work we focus on $\sigma_{\text{deg}}=1$, $\sigma_r > 0$ models as better reflections of future quantum networks which we refer to as the network class $\textsf{S}_{\sigma_r}$.

\section{Practical Routing and Performance-defined Criticality \label{sec:Prac_Routing}}

\subsection{Benchmarking Performance}
Consider a class of quantum networks $\textsf{N} = \{\mc{N}_i \}_{i}$ and a specific routing protocol $\mc{P}$. For any network $\mc{N} = (P,E) \in \textsf{N}$ drawn from this class, one can define the \text{average end-to-end rate} as that which is averaged over all possible end-user pairs in the network. We denote this set of end-users by 
\begin{equation}
\mc{I} \defeq \{ \bs{i}_j \}_{j} =  \{ \{ \bs{a}_j, \bs{b}_k \} \}_{j\neq k}.
\end{equation}
Then the average end-to-end rate takes the form,
\begin{equation}
\langle K \rangle_{\mc{N}|\mc{P}} \defeq \frac{1}{| \mc{I} |} \sum_{\bs{i} \in \mc{I}} K( \bs{i}, \mc{N} | \mc{P}) \label{eq:NetAvg}
\end{equation}
where $K( \bs{i}, \mc{N} | \mc{P})$ is the rate achieved between the end-user pair $\bs{i} = \{\bs{a}, \bs{b} \}$ given that they engage in the routing protocol $\mc{P}$ (given in Eq.~(\ref{eq:GenFlood})). In an $N$ node network there exist $C_N^2$ potential end-user pairs (where $C_n^k$ denotes the binomial coefficient), which may be far too large to consider explicitly. In practice, we will sample $L$ end-user pairs from $\mc{N}$ to approximate the average end-to-end rate,
$
\langle K \rangle_{\mc{N}|\mc{P}} \approx \frac{1}{L} \sum_{j=1}^L K( \bs{i}_j, \mc{N} | \mc{P}) \label{eq:NetAvg}
$
which can be computed along with statistical error considerations.

This quantity benchmarks the ability of quantum communication on a specific network. To assess the efficacy of a routing strategy more generally, we study the \textit{ensemble average end-to-end rate} sampled across the entire network class. Given that any network $\mc{N} \in \textsf{N}$ is drawn with equal probability, then the ensemble average rate is equal to
\begin{equation}
\langle K \rangle_{\textsf{N}|\mc{P}} \defeq \frac{1}{|\textsf{N}|} \sum_{\mc{N} \in \textsf{N}} \langle K \rangle_{\mc{N}|\mc{P}}. \label{eq:EnsAvg}
\end{equation}
For the network classes with which we are interested, $\textsf{N}$ can be incredibly large (or infinite). Hence, we are not be able to exactly compute Eq.~(\ref{eq:EnsAvg}) but instead compute an accurate approximation over a sample space of $M^\prime$ networks $\{\mc{N}_i\}_{i=1}^{L^\prime}$ such that
$
\langle K \rangle_{\textsf{N}|\mc{P}} \approx \frac{1}{L^\prime} \sum_{i=1}^{L^\prime} \langle K \rangle_{\mc{N}_i|\mc{P}} \label{eq:EnsAppx}.
$
This approximation can be made sufficiently accurate by taking enough end-user samples $L$ and network class samples $L^\prime$.

These concepts can be immediately translated in the context of end-to-end capacities rather than rates. Indeed, one can readily define a specific end-to-end capacity $\mc{C}( \bs{i}, \mc{N} | \mc{P})$, an average $\langle \mc{C} \rangle_{\mc{N}|\mc{P}}$ and ensemble average $\langle \mc{C} \rangle_{\textsf{N}|\mc{P}}$. The only difference is the description of the point-to-point rates throughout the network; when network links are considered to operate at their capacity, then we are studying the end-to-end capacities. Otherwise, they are sub-optimal end-to-end rates. 

\subsection{Benchmarking Efficiency}

Routing protocols should not only be benchmarked with respect to performance, but also with respect to efficiency, i.e.~what proportion of network resources is required to guarantee a particular level of performance?
Therefore, it is useful to define a measure which we call the \textit{routing consumption}, which quantifies the proportion of network edges used via a given routing protocol. Given a network $\mc{N} = (P,E)$, an end-user pair $\bs{i} = \{\bs{a},\bs{b}\}$ and a routing protocol $\mc{P}$, the routing consumption $\tilde{E}$ is the fraction of network edges required to perform communication,
\begin{equation}
\tilde{E}(\bs{i},\mc{N}|\mc{P}) \defeq \frac{ |E_{\mc{P}}(\bs{i},\mc{N}) | }{|E|},
\end{equation}
where $|E_{\mc{P}}(\bs{i},\mc{N})|$ is the number of edges in the routing edge set connecting the end-user pair. 
We can then define an average end-to-end routing consumption by averaging over end-user pairs
$
\langle\tilde{E}\rangle_{\mc{N}|\mc{P}} \defeq \frac{1}{|\mc{I}|} \sum_{\bs{i}\in\mc{I}} \tilde{E}(\bs{i},\mc{N}|\mc{P}),
$
which of course extends to an ensemble average over a class of networks 
$
\langle\tilde{E}\rangle_{\textsf{N}|\mc{P}} \defeq \frac{1}{|\textsf{N}|} \sum_{\mc{N}\in\textsf{N}} \langle\tilde{E}\rangle_{\mc{N}|\mc{P}},
$
following the notation convention used for end-to-end rates. As before, these quantities are estimated by sampling from the set of end-user pairs and the network class.

The routing consumption is a useful measure of how resource efficient a network protocol is at achieving its rates. A protocol which can attain high-rates with a low routing consumption is clearly a desirable strategy. Hence, a tradeoff exists between end-to-end rates and routing consumption. While we know flooding to be rate optimal, it will always satisfy $\langle{\tilde{E}\rangle_{\textsf{N}|\mc{P}_{\text{fl}}} = 1}$ because it utilises all network edges; hence it is most likely sub-optimal from the perspective of routing consumption. For single-path protocols we can typically expect $\langle \tilde{E}\rangle_{\textsf{N}|\mc{P}_{\text{sp}}} \ll 1$, but its rates may be poor. It is most interesting to inspect the behaviour of $\langle \tilde{E}\rangle_{\textsf{N}|\mc{P}_{\text{mp}}}$ as understanding the relationship between end-to-end rate, resource consumption and practical multi-path routing is least understood.

\subsection{Practical Routing}

In practice, one must be able to deploy a routing algorithm which informs network nodes how to interact and establish communication between the end-users. This algorithm characterises the network protocol, for which there are some intuitive, practical objectives. The algorithm should:
\begin{enumerate}
\item Identify an end-to-end path (or paths) which maximize the rate between users. Alternatively, the protocol should surpass a target rate requirement.
\item Minimize the network resources needed to achieve said end-to-end rate, i.e.~minimize the necessary network nodal density as well as the number of links/nodes which participate in communication.
\item Be efficient enough to run and facilitate routing without impeding the end-to-end rate. 
\end{enumerate}
It is extremely important that each of these points are considered for a routing strategy to be deemed practical. A easily executed protocol is useless if it identifies poor routes, while a protocol which establishes high rate routes very slowly is equally undesirable.

Single-path routing is the principal mechanism for classical communications, fundamentally achieved by Dijkstra's algorithm (DA) \cite{Dijkstra59}. Generally, this is a greedy algorithm for single-path route optimisation according to a path defined \textit{cost function}, which measures a property of end-to-end routes that one wishes to optimise, e.g.~minimisation of path length or maximisation of bottleneck rate. The challenges which extends from the point-to-point rate limitations of quantum communications means that rate optimisation is the salient task. This is also known solving the widest path problem. For a network $\mc{N} = (P,E)$, DA locates the widest path efficiently in run-time $\mc{O}(|E| + |P|\log_2 |P|)$. Given a single-path protocol $\mc{P}_{\text{sp}}$ and an end-user pair, the optimal end-to-end rate is given by
\begin{equation}
{K} (\bs{i}, \mc{N} | \mc{P}_{\text{sp}}) \defeq \max_{\omega} \min_{(\bs{x},\bs{y})\in \omega} {K}_{\bs{xy}} \label{eq:SPR},
\end{equation}
where the maximisation is performed over all possible end-to-end routes~\cite{Note4}.

%\footnote{This is equivalent to specifying Eq.~(\ref{eq:GenFlood}) with the forwarding probability distribution to 
%${\{ q_{\bs{xy}} = 1 \}_{(\bs{x},\bs{y})\in \omega^{\star}}} \cup \{ q_{\bs{xy}} = 0 \}_{(\bs{x},\bs{y}) \notin \omega^{\star}}$}.

To mitigate single-path performance limitations, multi-path routing emerges as good solution. 
Flooding represents a useful paradigm for benchmarking optimal network performance, but it is not practical in many real world settings. Utilizing an entire network to facilitate one end-user pair renders the network useless for \text{any other} set of users to communicate simultaneously. In reality, we want to enable the concurrent use of a network for many users. 

A basic approach to building high-rate multi-paths is to perform an iterative form of DA. This is a modified algorithm which locates multiple end-to-end routes which are either edge-disjoint (no two paths share the same edge) or node-disjoint (no two paths visit the same node). Both of these versions involve executing a Dijkstra search, followed by a network modification where edges included in previous paths (or in the neighbourhoods of nodes in previous paths) are not included in the next search. Locating $M$ end-to-end paths requires $M$ executions of DA, leading to an increased time complexity $\mc{O}( M (|E| + |P|\log_2 |P|))$. This a costly scaling factor which does not lend well to deployability in large-scale networks. 

\subsection{Efficient Multi-Path Routing}

Fortunately, there are ongoing developments in the study of fast algorithms for multi-path routing \cite{Abe15,Tapolcai19,Lemeshko19,Bernardo20,1SMDP1}. In particular, Lopez-Parajes \textit{et al.}~recent devised an efficient, centralised, \textit{one-shot} approach to multi-path routing through their Multiple Disjoint Path Algorithm (MDPAlg) \cite{1SMDP2}. They recognised that a single execution of DA observes much more information than it actually utilises. Given a source node $\bs{a}$, the standard DA focuses on building a minimum cost tree from which only the optimal end-to-end paths can be located from any other node $\bs{x} \in P \setminus \{\bs{a}\}$ in the network. It does this by storing a minimum cost set (stores the minimum cost associated with traversing from $\bs{a}$ to any node $\bs{x}$) and a parent node set (stores the parent node from which the minimum cost was obtained to $\bs{x}$). With these quantities, DA can then reconstruct a minimum cost path by backtracking from a target node $\bs{b}$ along the parent-node tree until $\bs{a}$ is reached.

Alternatively, the MDPAlg uses a \textit{cost matrix} to collect additional information on the \textit{aggregated} cost from $\bs{a}$ to any other node; not just the minimum cost. Whereas DA would discard information concerning sub-optimal paths, the MDPAlg stores such information in the cost matrix so that they may be used to construct additional end-to-end paths later. Furthermore, it does this through only a single search of the network rather than the many searches that may be required by an iterative Dijkstra approach. The cost matrix can then be used to reconstruct the optimal path, and many other paths between the source $\bs{a}$ and $\bs{b} \in P \setminus \{\bs{a}\}$. 

The original algorithm identifies multiple \textit{shortest paths} but can be modified to optimise alternative cost functions. In this work, we modify the MDPAlg in such a way that approximates rate optimisation by minimizing the \textit{inverse accumulated rate} over the course of an end-to-end route. We may call this variant \textit{IAR-MDPAlg}. We do not minimize the total route length, but rather the sum of its inverted rates. \CH{More precisely, we utilise the following optimisation ansatz to locate high-quality, low-resource intensive routes,
\begin{equation}
\omega^{\star} \defeq \argmin_{\omega \in \Omega} \sum_{(\bs{x},\bs{y}) \in \omega} (K_{\bs{xy}}^{-\eta} + \epsilon).
\end{equation}
Here, $\omega^{\star}$ is an optimised route, $\Omega$ is the set of all possible paths between a given end-user pair. Meanwhile $\eta, \epsilon \in \mathbb{R}_0^{+}$ are real hyperparameters. This aligns with minimizing the sum of the inverse point-to-point rates along a path $\omega$, adding a penalty for channel usage $\epsilon$ associated with each link. It follows that $\eta$ controls the strength of penalizing the use of links with weak rates, while $\epsilon$ controls the penalty strength of resource consumption. Minimizing this cost function simultaneously locates a path which maximizes the sum of the point-to-point rates along the path while minimizing its length; \textit{indirectly} identifying a high-rate and edge-efficient route.}

In this way, the algorithm will (typically) identify routes with large bottleneck rates. While this approach is approximate, we show in this work that it is effective (for more details we refer the reader to Appendices~\ref{sec:GenDijk} and \ref{sec:MDPAlg}). Furthermore, we focus on the edge-disjoint format under the assumption that repeaters may have multiple quantum registers that operate concurrently to route quantum systems along different channels. The edge-disjoint version can locate more end-to-end paths than its node-disjoint counterpart, thus offering greater utility.

There are two potential versions of multi-path protocol that we may consider. To enhance rates, one may insist upon the use of $M>1$ end-to-end paths, where $M$ is fixed. This is an intuitive approach which we call fixed route number protocols, denoted by $\mc{P}_{\text{mdp}}^{M}$. However, a pair of end-users may be more interested in preserving resources granted that they possess a particular rate guarantee. It is easy to devise a protocol in which there is a rate requirement; via the IAR-MDPAlg, we use the least number of end-to-end routes required to achieve $R^{\star}$ bits per protocol use. Denoted by $\mc{P}_{\text{mdp}}^{R^\star}$, this is easily implemented strategy and reflects the quality of service principle in classical networks. 

It is important to note that the IAR-MDPAlg is not optimal and may not always match the performance of iterative Dijkstra (or, of course, flooding). However, we will see that any minor cost in performance is vastly outweighed by gains in efficiency, as the IAR-MDPAlg can construct end-to-end multi-paths orders of magnitude faster than iterative Dijkstra approaches (see Ref.~\cite{1SMDP2} for cost analyses). This makes it a much more practical method and allows us to perform numerical assessments that would not be possible otherwise. 
For a more detailed explanation of the algorithms used in this study, we refer the reader to Appendix \ref{sec:MDPAlg}.

\subsection{Criticality of Quantum Networks}

It is essential to pursue a rigorous and quantitative assessment of practicality with respect to quantum network routing, and allow us to understand the core network features which guarantee both performance and efficiency. To this end, the concept of \textit{network criticality} is vital. A critical transition is an abrupt regime shift in the behaviour of some complex system, such that the system transits from sub-to-super-critical with respect to a particular property. In the context of complex quantum networks, we have already discussed how critical transitions may occur with regards to robustness (connectivity) but it can be extended to notions of performance \cite{QuntaoRandQNets} or routing efficiency. A major contribution of our work is to introduce routing consumption-criticality, identifying network regimes within which end-to-end routing is efficient via practical routing.

We begin with performance-defined criticality; critical properties within a quantum network model which a guarantee a transition from unreliable rates  to consistent super-critical rates. A useful manifestation of performance-based criticality that can arise is end-to-end distance independent rates, i.e.~the spatial separation of users is independent of communication quality. This is especially important to quantum networks in order to overcome point-to-point limitations. 
Furthermore, these measures are more appropriate than purely connectivity defined quantities, since they make meaningful assessments of the efficacy of a network to perform quantum communications. \CH{This quantity was first introduced in \cite{QuntaoRandQNets} in the context of network flooding protocols $\mc{P}_{\text{fl}}$ and was used to illuminate critical insight for quantum network composed of bosonic pure-loss channels. In this work, we explore such critical densities relative to a broader range of routing protocols and link-layer descriptions}. 

\begin{defin} \emph{(Performance-defined Critical Density):} For a class of quantum networks $\textsf{\emph{N}}$, and a network routing protocol $\mc{P}$, we define $\rho_{\textsf{\emph{N}}|\mc{P}}^{\star}$ as 
the minimum nodal density required to guarantee an ensemble average rate of {${\langle R \rangle_{\emph{\textsf{N}}|\mc{P}} \geq 1}$} bit/protocol use.
\end{defin}

The critical nodal density $\rho_{\textsf{{N}}|\mc{P}}^{\star}$ is an extremely important measure since quantum repeaters are costly and they should be minimized as a resource in a future quantum internet. Investigations of this quantity with respect to flooding $\rho_{\textsf{{N}}|\mc{P}_{\textsf{fl}}}^{\star}$ and capacity achieving bosonic lossy networks have been carried out, which can be considered a lower-bound for all other critical densities. Indeed, minimizing this quantity over all protocols is equivalent to the flooding defined critical density,
\begin{equation}
\rho_{\textsf{{N}}}^{\star} \defeq  \min_{\mc{P}} \rho_{\textsf{{N}}|\mc{P}}^{\star}   = \rho_{\textsf{{N}}|\mc{P}_{\textsf{fl}}}^{\star} \leq \rho_{\textsf{{N}}|\mc{P}}^{\star}.
\end{equation}
Nonetheless, analogous studies have not been carried out for other routing protocols or networks.

The performance-based critical density is a valuable characteristic of a quantum network model, however it lacks insight to the resources required to achieve critical rates. Networks may be super-critical with respect to performance but may demand impractical resources to do so. To address this we can analyse efficiency regimes with respect to routing, defining a consumption-based critical density.

\begin{defin} \emph{(Consumption-defined Critical Density):} 
Consider a class of quantum networks $\textsf{\emph{N}}$, and let $\textsf{\emph{N}}(\rho) \subset \textsf{\emph{N}}$ be a subset of this class with nodal density $\rho$. For a network routing protocol $\mc{P}$, we define $\tilde{\rho}_{\textsf{\emph{N}}|\mc{P}}^{\star}$ as the density at which the routing consumption is maximized and after which it undergoes consistent decay. More precisely, 
{${\tilde{\rho}_{\textsf{\emph{N}}|\mc{P}}^{\star} \defeq  \argmax_{\rho}\> \langle \tilde{E}\rangle_{\textsf{\emph{N}}({\rho})|\mc{P}}}$},
such that 
$
\langle \tilde{E}\rangle_{\textsf{\emph{N}}(\rho) |\mc{P}} \leq \langle \tilde{E}\rangle_{\textsf{\emph{N}}({\tilde{\rho}_{\textsf{\emph{N}}}^{\star}}) |\mc{P}},~\forall \rho \geq \tilde{\rho}_{\textsf{\emph{N}}}^{\star}.
$
\label{defin:deltaCrit}
\end{defin}

Hence, $\tilde{\rho}_{\textsf{N}|\mc{P}}^\star$ represents a secondary critical measure which separates network classes into different efficiency regimes. Executing the same protocol, networks which possess a nodal density $\rho > \tilde{\rho}_{\textsf{N}|\mc{P}}^\star$ will consume a smaller fraction of network resources during routing. Furthermore, this efficiency increase as the density continues to grow. We may connect the concepts of performance and routing consumption within the following definition.

\begin{defin} \emph{($\delta$-Critical routing consumption):} 
Consider a class of quantum networks $\textsf{\emph{N}}$. The critical routing consumption $\langle \tilde{E}\rangle_{\emph{\textsf{N}}}^{\star}$ is the minimum ensemble average fraction of network edges that must engage in end-to-end routing granted the ensemble average rate is $\langle R \rangle_{\emph{\textsf{N}}|\mc{P}} \geq \delta$ bits per network use, given that $\delta \leq \langle R \rangle_{\emph{\textsf{N}}|\mc{P}_{\text{\emph{fl}}}}$. More precisely, the critical routing consumption is
\begin{equation}
\langle \tilde{E}\rangle_{\emph{\textsf{N}},\delta}^{\star} \defeq  \min_{\mc{P} : \langle R \rangle_{\emph{\textsf{N}}|\mc{P}} \geq \delta} \langle \tilde{E}\rangle_{\textsf{\emph{N}}|\mc{P}} .
\label{eq:CritRC}
\end{equation}
\label{defin:deltaCrit}
\end{defin}

The definition in Eq.~(\ref{eq:CritRC}) can be made intuitive in the following way:
There exists a critical routing consumption achieved by an end-to-end protocol which can promise $\delta$ bits per network use on average, such that $\delta$ is an attainable network rate. Further to this performance guarantee, the protocol minimizes the ensemble average fraction of network edges used, i.e.~on average, it is the most efficient routing mechanism which promises the target rate. 
It is by no means clear how to determine the optimal protocol. However, the tools developed in this paper can effectively bound the critical routing consumption via practical routing schemes. In fact, we may write
\begin{equation}
 \langle \tilde{E}\rangle_{{\textsf{N}}|\mc{P}_{\text{sp}}} \leq \langle \tilde{E}\rangle_{{\textsf{N}},\delta}^{\star}  \leq \langle \tilde{E}\rangle_{{\textsf{N}}|\mc{P}_{\text{mp}}}. \label{eq:CRC_B}
\end{equation}
The optimal single-path protocol can always be a lower-bound, which is saturated iff an ensemble average $\delta$ bits per network use is achievable via single-path routing. An appropriate multi-path strategy generates the upper-bound. Indeed, there will always exist a multi-path strategy capable of this since one can employ a flooding protocol in the worst case (least efficient) scenario.

\begin{figure*}[t!]
\includegraphics[width=0.495\linewidth]{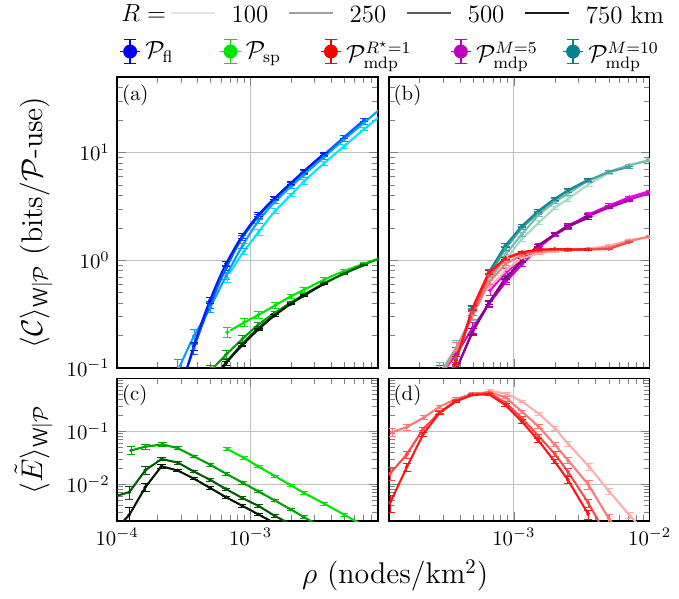}
\includegraphics[width=0.475\linewidth]{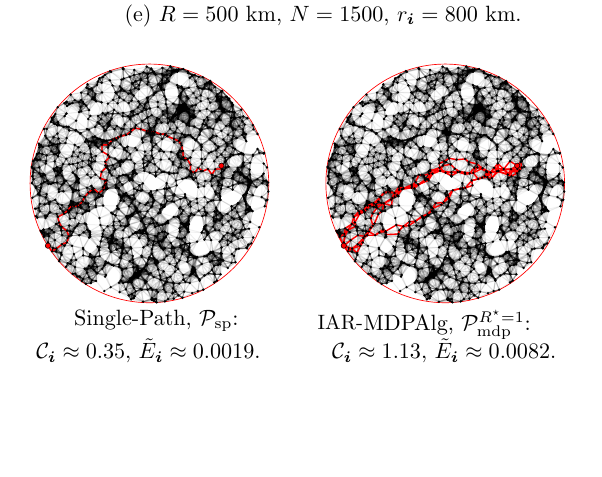}
\caption{Relationships between performance, routing consumption and nodal density in bosonic thermal-loss Waxman networks. The network link layer in all cases is described by the upper-bound capacity distribution $\mc{K}_u$ from Eq.~(\ref{eq:RD_u}) and model parameters $(r_0, \beta) = (100,1)$. Panels (a)-(b) plot the ensemble average end-to-end capacities achieved by different routing protocols and a number of network radii, $R$ (each protocol and network radius is colour coded with the routing protocol and $R$ legends). Panels (c)-(d) show the ensemble average routing consumption of (c) single path routing and (d) $\mc{P}_{\text{mdp}}^{R^{\star}=1}$ routing via the IAR-MDPAlg. 
Panel (e) depicts achievable end-to-end routes on an example network (according to the parameters listed) using $\mc{P}_{\text{sp}}$ and $\mc{P}_{\text{mdp}}^{R^{\star}=1}$ for users separated by $r_{\bs{i}} \approx 800$ km. Black edges in the networks identify unused edges, while red edges are those which engage in the routing protocol.}
\label{fig:WaxmanPerfs}
\end{figure*}

\section{Numerical Results}

\subsection{Benchmarking Waxman Fiber Networks}

In Fig.~\ref{fig:WaxmanPerfs} we present relationships between nodal density, routing consumption and end-to-end capacities on bosonic thermal-loss Waxman networks, with an upper-bounding capacity distribution $\mc{K}_{u}$ (lower-bounding capacity distribution $\mc{K}_l$). Each network edge represents an optical-fiber link of loss rate $0.2 \text{ dB/km}$ and environmental thermal noise $\bar{n} = 1/500$ connecting ideal transmitter/detectors. The Waxman parameters are set as $r_0,\beta = 100,1$ ($r_0,\beta = 63,1$), where the decay parameter corresponds to the approximate distance at which the single-edge capacity upper-bound collapses to zero $\sim 100$ km ($\sim 63$ km). Since we are considering end-to-end \textit{capacities}, results concerning these networks offer universal benchmarks for any fiber-based Waxman network. 

Figs.~\ref{fig:WaxmanPerfs}(a)-(b) depict the ensemble average end-to-end capacities with respect to nodal density for a number of routing strategies. In Ref.~\cite{QuntaoRandQNets}, the authors identified a performance-based critical nodal density of ${\rho_{\textsf{N}}^{\star} \approx 4.25\times 10^{-4}}$ nodes/km${}^2$ associated with flooding and bosonic pure-loss networks. The consideration of environmental thermal-noise naturally increases the critical density such that {$\rho_{\textsf{N}}^\star \approx {7.35 \times 10^{-4}}$}  ($1.04 \times 10^{-3}$) nodes/km${}^2$ (the density in parentheses refers to the lower-bounding capacity distribution). While significant, this shift remains relatively optimistic since the considered thermal-loss link layer does not consider additional experimental sources of thermal noise in an effort to remain protocol agnostic. 

Critical densities with respect to flooding protocols are inherently optimistic, due to the unrealistic resource demand. This optimism is emphasised when one considers the utility of single-path routing. Fig.~\ref{fig:WaxmanPerfs}(a) clearly illustrates the intense demands on the network density required for single-path routing to reach criticality, and promise effective end-to-end rates. We find the single-path critical nodal density to be approximately {${8.52 \times 10^{-3}}$} ($8.86 \times 10^{-3}$) nodes/km${}^2$, which is an order of magnitude larger than that predicted by flooding. On large-scales, a stark increase of this magnitude has significant ramifications for the cost and deployability of quantum networks. Therein lies the necessity for efficient multi-path routing protocols. Fig.~\ref{fig:WaxmanPerfs}(b) illustrates the efficacy of IAR-MDPAlg based routing, showing that flooding is not necessary to preserve lower critical densities. Each variant of the multi-path protocol is able to recover a critical density close to that offered by flooding, reinforcing the notion that high-rates can be guaranteed with realistic resources. 

Further evidence of this is gathered in Figs.~\ref{fig:WaxmanPerfs}(c)-(d) which plot the routing consumption of the protocols $\mc{P}_{\text{sp}}$ and $\mc{P}_{\text{mdp}}^{R^{\star}=1}$ respectively. While single-path protocols are expected to achieve very low routing consumptions, the same expectation is not necessarily held for multi-path routing. However, the protocol $\mc{P}_{\text{mdp}}^{R^{\star}=1}$ can obtain a critical density of approximately ${8.15\times 10^{-4}}$  ($1.21\times 10^{-3}$) nodes/km${}^2$, while maintaining a average maximum routing consumption of approximately $0.44$ ($0.36$). That is, no more than half of the network edges are ever required to participate in end-to-end routing. Even before single-path routing is a plausible option, the routing consumption of $\mc{P}_{\text{mdp}}^{R^{\star}=1}$ undergoes a rapid decay as the nodal density passes approximately $6.59\times 10^{-4}$ ($1.15\times 10^{-3}$) nodes/km${}^2$. For networks with nodal density $\rho \geq 3\times10^{-3}$, this protocol can guarantee super-critical rates while consuming less than $5\%$ of the network edges during routing. 

In this way, we can use the multi-path protocol $\mc{P}_{\text{mdp}}^{R^{\star}=1}$ to upper-bound the $\delta=1$ critical routing consumption as defined in Eq.~(\ref{eq:CRC_B}), i.e.~for bosonic thermal-loss networks we can write an upper-bound on the minimum network fraction required to guarantee an ensemble average of 1 bit/protocol-use, $\langle \tilde{E}\rangle_{{\textsf{W}},1}^{\star} \lesssim 0.44$. 
This places an efficiency bound on the optimal end-to-end network protocol, which is significantly more informative than flooding. 

Minimizing edge usage while guaranteeing high rates is vital to scalability in large quantum networks with many users. Crucially, in the absence of effective single-path routing, efficient multi-path strategies exist that preserve network resources while obtaining reliable, super-critical end-to-end rates. The importance of this result is emphasized when more realistic links are considered, as those described by practical QKD protocols (and beyond).

\subsection{Practical Link Layers and Network Phases}

Performing the previous analyses of end-to-end rates, routing consumption, and nodal density for different link layers, we can build a comprehensive picture of the efficacy of quantum communication networks in different scenarios. As such, we can establish relationships between different critical network properties which give rise to key \textit{network phases}; sub-classes of Waxman networks for which the behaviour of end-to-end quantum communication have similar characteristics. Here, phases are defined through statistical analyses of the properties deemed most important to end-to-end communication; connectivity, performance, and routing efficiency. 
Identifying what these phases are and where they fall within meaningful density ranges can help us to understand the realistic needs of quantum networking. 

\begin{figure*}[t!]
\hspace{-1.25cm} (a) Waxman Phase Transitions, \hspace{5.25cm} (b) Phase Descriptions.\\
\includegraphics[width=0.95\linewidth]{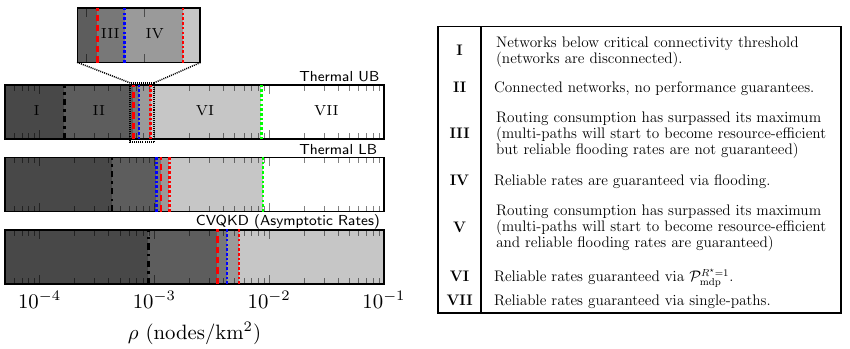}
\caption{Waxman quantum network phase characterisations with respect to link layer descriptions and nodal density. Panel (a) outlines connectivity, consumption and performance based critical densities with respect to networks composed of different link layers. These transitions give rise to network phases in which we can expect particular properties. These phases are labelled on the density diagram, while Panel (b) summarises and describes their implications for quantum networking. Phases \CH{III} and \CH{V} describe similar network properties, but differ as to whether the maximum routing consumption is surpassed before or after the flooding-based performance transition. Note that Phase \CH{III} appears in networks described by the thermal upper-bound capacity distribution, while Phase \CH{V} appears in the lower-bound capacity distribution. }
\label{fig:PhaseTrans}
\end{figure*}

Fig.~\ref{fig:PhaseTrans} summarises the Waxman quantum network phases defined in this study for a number of link layer models: bosonic thermal-loss capacity distributions, and asymptotic CV-QKD rate distributions. Here, we have identified six key network phases, explicitly defined in Fig.~\ref{fig:PhaseTrans}(b), ranging from Phase I networks with no connectivity guarantees, to Phase \CH{VII} networks which promise super-critical performance via single-path routing. In between, there exists a spectrum of phases corresponding to critical changes in the connectivity, routing consumption and performance guarantees.

\CH{These phases have been identified by computing and comparing various critical densities relative for each link layer model. Let us define these phases more precisely: 
\begin{itemize}
    \item \textbf{I}: Densities fall below the giant-component transition density, $\rho < \rho_{\text{G}}^{\star}$.
    \item \textbf{II}: Densities are greater than the giant-component transition density, but smaller than the flooding-based critical density, $\rho_{\text{G}}^{\star} \leq \rho \leq \rho_{\textsf{W}|\mc{P}_{\text{fl}}}^{\star}$.
    \item \textbf{III}: Densities are greater than the consumption-defined critical density, but smaller than the flooding-based critical density, ${\tilde{\rho}_{{\textsf{W}|\mc{P}_{\text{fl}}}}^{\star} \leq \rho \leq \rho_{\textsf{W}|\mc{P}_{\text{fl}}}^{\star}}$. Not all link-layer models will inhabit this phase.
    \item \textbf{IV}: Densities are greater than the flooding-based critical density, $\rho \geq {\rho}_{{\textsf{W}|\mc{P}_{\text{fl}}}}^{\star}$.
    \item \textbf{V}: Densities are greater than the flooding-based critical density and greater than the consumption-defined critical density $\rho \geq \tilde{\rho}_{{\textsf{W}|\mc{P}_{\text{fl}}}}^{\star} \geq {\rho}_{{\textsf{W}|\mc{P}_{\text{fl}}}}^{\star}$. Not all link-layer models will inhabit this phase.
    \item \textbf{VI}: Densities are greater than the $\mc{P}_{\text{mdp}}^{{R}^{\star}=1}$-based critical density, $\rho \geq {\rho}_{{\textsf{W}|\mc{P}_{\text{mdp}}^{R^{\star}=1}}}^{\star}$.
    \item \textbf{VII}: Densities are greater than the single-path-based critical density, $\rho \geq {\rho}_{\textsf{W}|\mc{P}_{\text{sp}}}^{\star}$.
\end{itemize}
Fig.~\ref{fig:PhaseTrans}(b) summarises these phases and interprets them from a practical perspective.
}

The most desirable network phase is naturally Phase \CH{VII}, in which single-path routing is sufficient to perform reliable quantum communication. Unfortunately, our primary takeaway is that the nodal densities necessary to inhabit Phase \CH{VII} are very large, and become impractical as point-to-point link layer descriptions become more realistic. Even when considering fully trusted QKD networks described using \text{asymptotic} secret key-rates, the nodal densities required to reach Phase \CH{VII} are more than $10^{-1}$ nodes/km${}^2$. For example, the number of nodes needed to deploy a fully trusted QKD network that spans the surface area of  Europe ($\sim 10^{7}$ km${}^2$) would be on the order of millions. 

Fortunately, efficient multi-path routing strategies can resurrect the utility of lower density networks, which achieve reliable rates. Networks which occupy Phase \CH{VI} are super-critical provided that they employ the protocol $\mc{P}_{\text{mdp}}^{{R}^{\star}=1}$, or better. As shown in Fig.~\ref{fig:PhaseTrans} for each considered link layer, Phase \CH{VI} occupies a more practical nodal density region (around an order of magnitude improvement) for which practical multi-path strategies can promise strong rates using efficient protocols; not just flooding.

An important observation that we wish to reiterate is the existence of Phases II and \CH{VI}. Phase II portrays the resource gap between connectivity guarantees and the most optimistic performance guarantee (via flooding). The existence of this gap makes it clear that the design and assessment of quantum networks \textit{cannot} solely focus on connectivity analyses, as such promises are not enough to guarantee useful quantum communication. Analogously, Phase \CH{VI} identifies a gap between the resources required by flooding and practical multi-path protocols for critical rates. Closing the gap posed between Phase \CH{VI} and \CH{VII} while minimizing routing consumption is the goal of any multi-path strategy.

These insights emphasise the need for explicit analyses of end-to-end performance and reiterate the point that routing in quantum networks cannot na\"{i}vely follow its classical counterpart. Multi-path routing techniques do not just represent a means of boosting rates but establish a pertinent method of reducing the resource demands of practical quantum network development. \CH{We reiterate that these phases are specifically derived for the Waxman network model, which is defined by its relationship between link-length and network connectivity. Nonetheless, the rate-distance tradeoff is a quintessential element of quantum communications; hence, these results provide an informative window into future quantum network properties.}

\subsection{Scale-Free Properties and Quantum Networks}
Scale-free architectures capture important features of real world networks beyond that of the Waxman model. As discussed in Section~\ref{sec:SFNets}, scale-free networks obey a power-law cumulative degree distribution. This realises a connectivity structure in which there exist a number of highly connected hubs to which many nodes of low degree are connected, giving rise to a lower average degree than Waxman models. Considering the network class $\textsf{S}_{\sigma_r}$, we can investigate the ramifications these connectivity features have on end-to-end routing and performance 

\begin{figure*}[t!]
\includegraphics[width=0.49\linewidth]{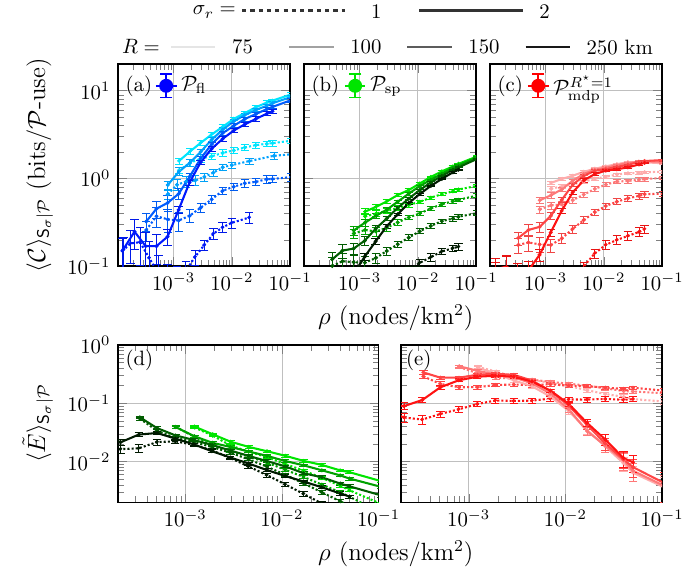}
\includegraphics[width=0.47\linewidth]{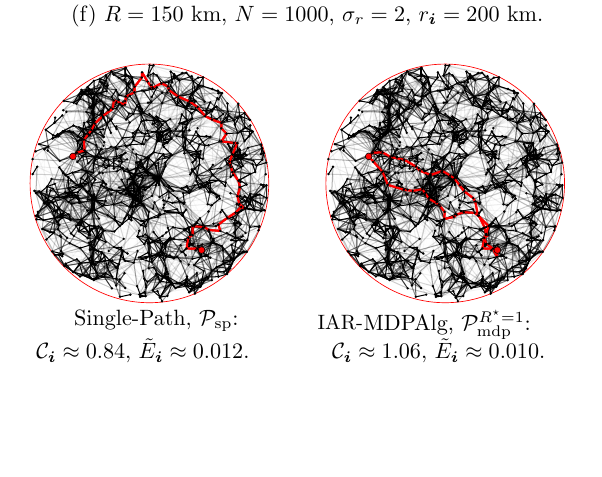}
\caption{Relationships between performance, routing consumption and nodal density in bosonic thermal-loss scale-free networks. Throughout all plots, the network link layer is described by the upper-bound capacity distribution $\mc{K}_u$ from Eq.~(\ref{eq:RD_u}) and we use the model parameters $(n_0, m, \sigma_{\text{deg}}) = (10,5,1)$. We also consider unique values for the scale-free model parameter $\sigma_r \in \{1,2\}$ which are distinguished in the legend. Panels (a)-(c) depict the ensemble average capacity with respect to nodal density for $\mc{P}_{\text{fl}}$, $\mc{P}_{\text{sp}}$ and $\mc{P}_{\text{mdp}}^{{R}^{\star}=1}$ routing respectively. Panels (d)-(e) plot the ensemble average routing consumption with respect to nodal density for $\mc{P}_{\text{sp}}$ and $\mc{P}_{\text{mdp}}^{{R}^{\star}=1}$ routing respectively. Each analysis is performed for a number of network radii $R$ listed in the legend. 
Panel (f) illustrates achievable end-to-end routes on an example network (under the parameters shown) using $\mc{P}_{\text{sp}}$ and $\mc{P}_{\text{mdp}}^{R^{\star}=1}$ for users separated by $r_{\bs{i}} \approx 200$ km. Black edges in the networks identify unused edges, while red edges are those which engage in the routing protocol.}
\label{fig:SFthermB}
\end{figure*}

Fig.~\ref{fig:SFthermB} displays analyses of bosonic thermal-loss scale-free networks with respect to nodal density and a number of routing protocols: Flooding, single-path, and $\mc{P}_{\text{mdp}}^{{R}^{\star}=1}$ routing using the edge-disjoint IAR-MDPAlg. We consider architectures generated with the parameter $\sigma_r  \in \{1,2\}$ so to vary the awareness of link-length during network creation and consequently its ``strictness" with respect to scale-free behaviour. In Figs.~\ref{fig:SFthermB} (a)-(c) we study the ensemble average end-to-end capacities for each network model and routing strategy. It is clear for $\sigma_r = 1$ (in which scale-free behaviour is followed closely) that performance is restricted by the low average degree. At small network scales (e.g.~$R < 150$ km) and increasing density, the flooding capacity is able to reach reliable rates, but only at very high densities. Furthermore, Ref.~\cite{QuntaoRandQNets} showed that the optimal flooding capacity of such networks exponentially decays with respect to $R$ (regardless of the number of nodes). 

Practical routing strategies cannot do better, as displayed for both single-path routing and $\mc{P}_{\text{mdp}}^{{R}^{\star}=1}$ routing. Ultimately, the connectivity structure of scale-free networks $\mc{N}\in\textsf{S}_{1}$ display a poor aptitude for multi-path routing. The utilisation of multiple routes is only helpful if many additional routes can be found, and the existence of high-degree hubs limits this possibility. User nodes will typically only connect to a single hub, limiting the ability to reinforce the end-to-end path set. This limitation can be seen in the ensemble average routing consumption of the $\mc{P}_{\text{mdp}}^{{R}^{\star}=1}$ protocol, which remains relatively constant with respect to nodal density. This means that there is little correlation between increasing density and the number of effective end-to-end routes between user nodes.

Breaking from strict scale-freedom, we see that effective performance \textit{can} re-emerge within the network class $\textsf{S}_{2}$. With an increased connection probability dependence on link-length, the ensemble average capacity and routing consumption follow similar trends to the Waxman model: The exponential decay of the flooding capacity with respect to network scale is subdued, and end-to-end capacities of $\langle \mc{C} \rangle_{\textsf{S}_{2}|\mc{P}} \geq 1$ bit/protocol-use can be guaranteed for each routing protocol (within reasonable nodal density ranges). Furthermore, the multi-path routing consumption reliably decays with increasing density, implying that an applicability for multi-path routing can be reestablished.

\section{Discussion\label{sec:Conc}}

In this work, we have investigated the capabilities of quantum communication networks under practical routing strategies, in an effort to gain insight to realistic requirements of future quantum networks. With the goal of providing a comprehensive study of the performance and feasibility of quantum optical fiber-networks, we have combined theory of quantum communication with that from random network models and network routing algorithms. Our work departs from previous developments in this domain by considering realistic link layer descriptions (capacities and secret-key rates over bosonic thermal-loss channels) alongside practical routing protocols (efficient multi-path and single-path routing methods). These assessments focus on the crucial Waxman and scale-free classes of random quantum networks. 

Through our statistical analyses, we reveal vital insights into the practical requirements of critical behaviour in quantum networks, where criticality may be defined with respect to connectivity, performance and routing efficiency. For quantum Waxman networks, a network phase structure is introduced with respect to nodal density; identifying ranges of nodal densities within which networks can be classified as sub- or super-critical with respect to these properties. We show that for increasingly realistic link-layer descriptions, the existence of a performance super-critical phase with respect to single-path routing becomes more and more difficult within practical density ranges. Nonetheless, multi-path routing strategies can re-establish practical density ranges and super-critical rates, reiterating the need for the development of multi-path routing methods in quantum networks. 

It is important to note that the network phases identified are classifications which emerge naturally from the heuristics considered in this work but are by no means definitive. With future, increasingly sophisticated investigations we expect richer and more detailed phases to be unveiled which provide insight and guidance for quantum network design.

Our findings reiterate the challenging relationship between scale-free networks and quantum communication. While scale-free architectures such as Yook's model \cite{Yook} offer useful insight into the structure of the classical internet, they do not find an analogy with reliable quantum fiber networks. Adherence to scale-freedom in large-scale quantum networks results in a poor communication quality regardless of routing strategy. By altering Yook's model to generate networks with an increased sensitivity to link length, scale-freedom begins to break while the ability to perform quantum communication strengthens. These results stress the resounding differences between the classical and quantum internet: Quantum networks need to be \textit{consistently} well connected to overcome point-to-point limitations, and must be strictly engineered with nodal density and link length in mind. This is a crucial take-home message that our work has established.

While our results focus on the efficacy of fiber networks, it is well motivated that the quantum internet will exploit satellite links and inter-satellite networks to reinforce long-range communication. This is an area of serious interest and a future investigative path for extending this research, investigating the interplay of realistic, ground-based networks interacting with satellite-based infrastructure. Furthermore, the study of ground-based free-space networks in both long-range (inter-metropolitan) and mobile (metropolitan) settings is of immediate interest. 

Understanding how many end-user pairs may simultaneously perform quantum communication across realistic networks is of paramount importance. Indeed, the trade-off between network architecture, end-to-end rates, and the number of end-users is still poorly understood. While the study of end-to-end routing consumption is a useful step in this direction, greater progress must be made in order to develop practical routing strategies for the quantum internet. \SP{In particular, future work should extend these results to the case of entanglement distribution via imperfect quantum repeaters, where decoherence, operational errors, and finite memory lifetimes will further impact network performance and routing efficiency.}

\acknowledgements
This work was supported by the EPSRC (Grant No. EP/R513386/1) and UKRI through the Integrated Quantum Networks (IQN) Research Hub (Grant No. EP/Z533208/1). C.H thanks Alasdair I.~Fletcher for helpful comments and discussions.

\appendix
\section{Asymptotic CV-QKD Secret Key-Rates \label{sec:KeyRates}}

In the main-text we consider a realistic link-layer for fully trusted QKD networks, in which all point-to-point links utilise a Gaussian-modulated, coherent-state-based CV-QKD protocol and heterodyne detection \cite{PracKeyRates}. By assuming that the nodes are trusted, we can assume that each link utilises a point-to-point QKD protocol. Furthermore, we consider (\textit{i}) asymptotic secret key-rates in the asymptotic limit of many system exchanges (discounting the additional detrimental effects of parameter estimation or finite key-lengths) and (\textit{ii}) all imperfections in the trusted communicators' experimental setup are untrusted, i.e.~all photons lost along the channel \textit{and} at the receiver are leaked into Eve's environment. 

Let us be more precise: Consider a point-to-point fiber channel connecting Alice and Bob who employ the Gaussian modulated coherent-state CV-QKD protocol outlined in Ref.~\cite{PracKeyRates}. Let us denote the total channel loss as $\tau = \eta_{\text{eff}} \eta_{\text{ch}}$ where $\eta_{\text{ch}}$ is the usual fiber-channel loss and $\eta_{\text{eff}}$ is the detector efficiency (which we take to be $\eta_{\text{eff}} = 0.7$). Alice and Bob exchange $\bar{n}_T$ photons from the transmitter, but at the receiver there are $\bar{n}_R$ photons at the receiver, where 
\begin{equation}
\bar{n}_R \defeq \tau \bar{n}_T + \bar{n}, 
\end{equation}
such that $\bar{n} \defeq \eta_{\text{eff}} \bar{n}_{\text{bg}} + \bar{n}_{\text{ex}}$ accounts for additional photons due to inefficiencies, background noise and experimental setup noise. In this protocol, the primary contribution of setup noise is due to the local oscillator, which in the local-local oscillator (LLO) protocol contributes both phase and electronic errors (see Ref.~\cite{PracKeyRates} for more details).  

Using the protocol's entanglement-based representation, and assuming a coherent state modulation of $\mu$, Alice and Bob share the Gaussian state $\rho_{AB}$ after the thermal-loss channel with covariance matrix,
\begin{equation}
\bff{V}_{AB} \defeq \begin{pmatrix}
\mu	\bff{I} & \sqrt{\tau(\mu^2-1)} \\  \sqrt{\tau(\mu^2-1)} & b \bff{I}
\end{pmatrix}
\end{equation}
where $b \defeq \tau(\mu-1) + 2\bar{n} +1$. Given that $\nu_{\pm}$ denotes the symplectic eigenvalues of $\bff{V}_{AB}$, Eve's Holevo bound is given by
\begin{equation}
\chi_E \defeq h(\nu_+)  + h(\nu_-) - h\left(\mu - \frac{\tau(\mu^2-1)}{b+1}\right).
\end{equation}
where $h(x) \defeq \frac{x+1}{2} \log_2(\frac{x+1}{2}) - \frac{x-1}{2} \log_2(\frac{x-1}{2})$. Consequently, the secret key-rate is 
\begin{equation}
K_{\tau,\bar{n}} \defeq \beta I_{AB}^{\tau,\bar{n}} - \chi_E^{\tau,\bar{n}}
\end{equation}
where $I_{AB} \defeq \log_2\left(1+ \frac{\tau(\mu-1)}{2(\bar{n}+1)}\right)$ is the mutual information between Alice and Bob and $\beta \in [0,1]$ is the reconciliation efficiency which accounts for imperfect data-processing (which we approximate as $\beta = 0.95$).

\begin{figure}
\includegraphics[width=\linewidth]{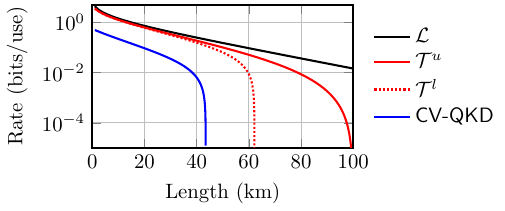}
\caption{Behaviour of point-to-point bosonic pure-loss capacity $\mc{L}$, thermal-loss capacity bounds $\mc{T}^{u/l}$ and asymptotic CV-QKD rate with respect to channel length. }
\label{fig:SE_Rates}
\end{figure}

This point-to-point rate can then be used to describe the communication rate along a single network edge, $K_{\bs{xy}} = K_{\tau_{\bs{xy}},\bar{n}_{\bs{xy}}}$ and provide an accurate link-layer described for trusted node QKD networks. 
Fig.~\ref{fig:SE_Rates} displays a comparison of the single-edge rates asymptotic CV-QKD rate, bosonic thermal-loss capacity bounds and the exact bosonic pure-loss capacity with respect to channel length.

\section{Network Cuts \label{sec:CutSet}}

An important graph-theoretic concept in the context of network performance is that of \textit{cuts} and \textit{cut-sets}. Consider a network $\mc{N}=(P,E)$ with two remote end-users $\bs{a}, \bs{b} \in P$. We define a cut $C$ as a bipartition of all network nodes $P$ into two disjoint subsets $(P_{\bs{a}}, P_{\bs{b}})$ such that the end-user nodes become completely disconnected, $\bs{a} \in P_{\bs{a}}$ and $\bs{b} \in P_{\bs{b}}$, where $P_{\bs{a}}\cap P_{\bs{b}} =\varnothing$. A cut $C$ generates an associated cut-set; a collection of network edges $\tilde{C}$ which when removed cause the partitioning,
\begin{equation}
\tilde{C} = \{ (\bs{x},\bs{y}) \in E~|~\bs{a} \in P_{\bs{a}}, \bs{b}\in P_{\bs{b}}\}.
\end{equation}
Under the action of a cut, a network is successfully partitioned
\begin{equation}
\mc{N} = (P,E) \xrightarrow{\text{Cut: $C$}} (P,E \setminus \tilde{C}) = ( P_{\bs{a}}\cup P_{\bs{b}}, E \setminus \tilde{C} ),
\end{equation}
so that there no longer exists a path between $\bs{a}$ and $\bs{b}$. Network cuts play a key role in the derivation of end-to-end network rates.

\section{Generalisations of Dijkstra's Algorithm\label{sec:GenDijk}}
It is possible to construct general versions of DA in which one optimises path-wise properties associated with end-to-end routes. It is well known that DA can be used to minimize path length (shortest path problem) or modified to maximize the bottleneck rate (widest path problem). But one can be more general and can define a global cost function $\mc{F}_{\omega}$ such that goal of DA is to find
\begin{equation}
\omega^{\star} = {\argX_{\omega}}~ \mc{F}_\omega,
\end{equation}
such that $X \in \{ \min,\max\}$. In this way, $\mc{F}_\omega$ may admit a more complex characterisation of routing value. 

Consider an $N$-node network $\mc{N} = (P,E)$. Let us define an $N$ element \textit{tentative path cost set} $T = \{T_{\bs{x}}\}_{\bs{x}\in P}$, which will be used to track the cost of routing throughout the search. This set will be initialised with a unique value for the source node $T_{\bs{s}} = \vep_{\text{init}}$, while all other nodes are initialised with a different value $T_{\bs{x}} = {\vep}_{\text{init}}^\prime$ for all $\bs{x} \in P\setminus \{\bs{s}\}$. Thus, $T$ takes the initial form
\begin{equation}
T = \{  {\vep}_{\text{init}}^\prime, \ldots, {\vep}_{\text{init}}^\prime, \underbrace{{\vep}_{\text{init}}}_{\text{source}},{\vep}_{\text{init}}^\prime, \ldots,  {\vep}_{\text{init}}^\prime\}.
\end{equation}
Along with this information, we also have access to point-to-point properties of each edge in the network. There may exist $m$ different types of properties for every edge $(\bs{x},\bs{y}) \in E$, which we denote via the set $\{ c_{\bs{xy}}^1, \ldots, c_{\bs{xy}}^m\}$.

The algorithm then operates greedily; starting at an initial source node $\bs{s}$, it traverses throughout the network hopping from node to node while keeping track of a tentative path cost from $\bs{s}$ to any other network node $\bs{x} \in P$.
Suppose that we have traversed from $\bs{s}$ to a node $\bs{x}$ and must decide whether to move to its neighbour $\bs{y}$ or not. Here, we must evaluate the tentative cost to the path if we make this move, since we should only hop to a subsequent node if it optimises the routing cost. The tentative path cost is evaluated via a \textit{tentative cost function} $F_{\omega}$ which admits the form
\begin{equation}
F_{\omega}^{\bs{s},{\bs{x}\rightarrow \bs{y}}}  = F_{\omega} ( T_{\bs{x}}, T_{\bs{y}}, \{ c_{\bs{xy}}^1, \ldots, c_{\bs{xy}}^m\}).
\end{equation}
That is, $F_{\omega}$ is a function of $T_{\bs{x}}$ the tentative path cost from $\bs{s} \rightarrow \bs{x}$, $T_{\bs{y}}$ the tentative path cost from $\bs{s} \rightarrow \bs{y}$ and the known point-to-point properties associated with moving along the edge $(\bs{x},\bs{y})\in E$. Note that $F_{\omega}$ is not equal to the global cost function $\mc{F}_{\omega}$, but they are inexorably tied. The tentative cost function is simply a translation of the global cost function into a form that can be employed within Dijkstra's algorithm. 

With the ability to evaluate the impact this hop has on the total path, the value $F_{\omega}^{\bs{s},{\bs{x}\rightarrow \bs{y}}}$ 
can be compared with the tentative path cost $T_{\bs{y}}$. If the goal is to minimize the path cost, then the search will hop to $\bs{y}$ iff $F_{\omega}^{\bs{s},{\bs{x}\rightarrow \bs{y}}} < T_{\bs{y}}$ i.e.~moving to this edge better minimizes the cost function. Contrarily, if the goal is to maximize the path cost function then the search will hop to $\bs{y}$ iff $F_{\omega}^{\bs{s},{\bs{x}\rightarrow \bs{y}}} > T_{\bs{y}}$. The algorithm follows these steps, evaluating, hopping and reevaluating the path cost until eventually the target node $\bs{t}$ is reached. Since the algorithm is greedy, it follows an optimal path throughout the network at all times. Once it arrives at a node, one can then be sure that the path followed has been the best one. 

Let us specify to a couple of examples. When minimizing the path length the global and tentative cost functions take the form
\begin{align}
\begin{aligned}
\mc{F}_{\omega} &= \sum_{(\bs{x},\bs{y}) \in \omega} l_{\bs{xy}},\\ 
F_{\omega}^{\bs{s},{\bs{x}\rightarrow \bs{y}}}  &=  T_{\bs{x}} + l_{\bs{xy}}, \label{eq:ShP}
\end{aligned}
\end{align}
where $l_{\bs{xy}}$ is a measure of point-to-point length of the edge $(\bs{x},\bs{y})\in E$. 
To minimize this quantity we use the initialisation values are $(\vep_{\text{init}}, \vep_{\text{init}}^{\prime}) = (0, \infty)$. 

For maximizing the bottleneck rate we define the global and tentative cost functions,
\begin{align}
\begin{aligned}
\mc{F}_{\omega} &= \min_{(\bs{x},\bs{y}) \in \omega} K_{\bs{xy}},\\
F_{\omega}^{\bs{s},{\bs{x}\rightarrow \bs{y}}}  &=  \max( T_{\bs{y}}, \min( T_{\bs{x}}, K_{\bs{xy}})), \label{eq:WiP}
\end{aligned}
\end{align}
where $K_{\bs{xy}}$ is the point-to-point rate of the edge ${(\bs{x},\bs{y})\in E}$. Here, we use $(\vep_{\text{init}}, \vep_{\text{init}}^{\prime}) = (\infty, -\infty)$.
\begin{figure}[t!]
\begin{algorithm}[H]
    \caption{General Dijkstra Algorithm}
    %\hspace*{\algorithmicindent} 
    \textbf{Inputs:} Network - $\mc{N}=(P,E)$, Source - $\bs{s}$, Target - $\bs{t}$,\\
    %\hspace*{\algorithmicindent}
     %\hspace*{\algorithmicindent} 
     Min/Max Functions - $(X,f_X) \in \{(\min, <) ,(\max,>)\}$.\\
     Tentative Cost Function - $F_{\omega}$, \\
    %\hspace*{\algorithmicindent} \textbf{Output} $\omega$ - Widest path.
    \vspace{3mm}
    \begin{algorithmic}[1] % The number tells where the line numbering should start
        \Procedure{MinCostPath}{$\bs{s},\bs{t},\mc{N}$}
	   \State Priority queue $Q = \{ \bs{x} \}_{\bs{x}\in P}$
	   \State Parent node set: $P = \{ \textsf{undef} \}_{\bs{x}\in P}$
	   \State Tentative cost set: $T = \{ \vep_{\text{init}}^{\prime} \}_{\bs{x}\in P}$ 
	   \State $T_{\bs{s}} \gets \vep_{\text{init}}$ 
	
	   \vspace{0.25cm}
	   \While{$|Q| > 0$} 
	   	\State $\bs{u} \gets \argX_{\bs{x}\in Q} T_{\bs{x}}$ (and remove $\bs{u}$ from $Q$) \vspace{1mm}
		\If{$\bs{u} \neq \bs{t}$}
			\For{\textbf{all} neighbours $\bs{v} \in Q$ of $\bs{u}$}
				\State $a \gets F_{\omega}(T_{\bs{v}}, T_{\bs{u}}, \{ c_{\bs{uv}}^{i}\}_{i=1}^m)$
				\If{$f_X(a,T_{\bs{v}})$ is true} 
					\State $T_{\bs{v}} \gets a$, $P_{\bs{v}} \gets \bs{u}$
					\State Reprioritise $Q$ wrt $T$
					\vspace{3mm}
				\EndIf
			\EndFor
		\EndIf
	\EndWhile
	\Return{ \textsc{ConstructPath}($P$, $T$)}
        \EndProcedure
    \end{algorithmic}
\end{algorithm}
\caption{Generalised Dijkstra's algorithm for end-to-end route optimisation with respect to a cost function $\mc{F}_{\omega}$. Any cost function has a tentative counterpart $F_\omega^{\bs{s},\bs{x}\rightarrow\bs{y}}$ which is used to evaluate movement throughout the network. The above pseudocode describes the network exploration phase which is followed by a \textsc{ConstructPath} subroutine which simply back tracks from the target node $\bs{t}$ to $\bs{s}$ using the constructed tentative cost and parent node sets.
}
\label{fig:GenDA}
\end{figure}

\section{The Multiple Disjoint Paths Algorithm (MDPAlg) and its variant\label{sec:MDPAlg}}
\subsection{MDPAlg}
As discussed in the main text, the MDPAlg offers an efficient means of determining multiple disjoint paths from a source $\bs{s}$ to a target $\bs{t}$ (or targets $\{\bs{t}_i\}_i$) within a network. While DA evaluates and stores the minimum cost to travel between the source and target nodes, the MDPAlg acquires additional information about the \textit{accumulated} cost of traversing from a source to target node through any of its neighbours \cite{1SMDP2}. This is enormously useful, and provides a mechanism for identifying many cost efficient routes between the source and target. 

Much like DA, the algorithm is split into two phases, (\textit{i}) network exploration and (\textit{ii}) path reconstruction. Throughout network exploration the algorithm proceeds similarly to DA in which the tentative cost matrix $T$ is constructed. The primary difference between DA and the MDPAlg is found on line 11 of Fig.~\ref{fig:GenDA}: Even when the computed tentative cost $a$ is not considered optimal, in the MDPAlg it is stored as an off diagonal element of $T_{\bs{uv}}$, describing the accumulated cost of travelling from node $\bs{s}$ to $\bs{v}$ \textit{via} $\bs{u}$. 

This additional information is then used in the path reconstruction phase to identify many end-to-end routes. The reconstruction is sequential and straightforward: Starting at a target node $\bs{t}$ a path is built by moving to the neighbour which incurs the lowest cost in the tentative cost matrix, $T$, until the source node is reached. On the first path reconstruction, this is simple and the minimum cost path $\omega_1$ is produced, which adds edges to the routing edge set ($E_{\omega} = \omega_1$). In order to construct subsequent disjoint paths, one must then enforce disjointedness by restricting the use of any edges which were used previously (for link disjointedness). In other words, future reconstructions will ignore the cost matrix elements $T_{\bs{xy}}$ for $(\bs{x},\bs{y}) \in E_\omega$. Repeated this process, path reconstruction may occur by moving from the target $\bs{t}$ to its \textit{next} minimum cost neighbour, and so on until $\bs{s}$ is met again. 

Through this sequential process of path building coupled with edge restriction, a number of edge-disjoint paths can be built which have favourable properties. The node-disjoint variant of the MDPAlg can be easily achieved through a small modification; instead of only restricting previously used edges from subsequent path reconstructions, one must restrict the use of any edges connected to nodes used in the previous routes. 
%For more details on this algorithm and its implementation, see Ref.~\cite{1SMDP2} where it was developed. 

\subsection{Rate Maximisation and IAR-MDPAlg}

The MDPAlg was introduced with the intention of identifying multiple disjoint \textit{shortest} paths which minimize the cumulative cost of edge-weights along an end-to-end route. 
Consequently, it is not immediately clear how the MDPAlg can be translated for the purposes of rate maximisation. One might assume that since there exists a variant of DA for this purpose (the widest path algorithm outlined in Appendix~\ref{sec:GenDijk}) then it should be able to modify the MDPAlg in an identical way. 

Unfortunately, this is not the case and we must be careful in building this variant (IAR-MDPAlg). The widest path version of DA locates an end-to-end route which maximizes a bottleneck rate between source and target nodes. It does so using via the generalised DA and the cost functions in Eq.~(\ref{eq:WiP}). 

Now let us consider the scenario in which we possess a source node $\bs{s}$, target node $\bs{t}$ and wish to find multiple end-to-end routes which maximize their bottleneck rates, e.g.~in the vein of iterative Dijkstra. We want to employ the more efficient MDPAlg to do this, in which only a single Dijkstra search is necessary, followed by path reconstruction. 
In this context, what information would a tentative cost matrix, $T$, collect? The diagonal element $T_{\bs{xx}}$ would contain the minimum cost associated with routing between $\bs{s}$ and $\bs{x}$, i.e.~the bottleneck rate along the optimal route. Meanwhile, the off-diagonal quantities $T_{\bs{xy}}$ indicate the potentially sub-optimal cost incurred when routing between $\bs{x}$ and $\bs{s}$ through the intermediate node $\bs{y}$. 

On the first iteration of path reconstruction, it is easy to identify the minimum cost path $\omega_1$ by backtracking a route to its most favourable neighbour node $\bs{n}_1$ and repeating this process according to subsequent nodes and the diagonal elements of $T$. Since we wish to construct edge-disjoint paths, any edge contained within $\omega_1$ is then rendered inaccessible by future routes.
Following $\omega_1$, we wish to construct a second end-to-end path using information from the cost matrix. This is initiated by hopping from the target $\bs{t}$ to its next more favourable neighbour, which we label $\bs{n}_2$. The cost matrix has collected seemingly valuable information about how to best traverse between $\bs{s}$ and $\bs{t}$ through $\bs{n}_2$. 

This is where our issue begins to emerge: What if the best path between $\bs{n}_2$ and $\bs{s}$ shares edges with $\omega_1$ (the first minimum cost path)? Widest paths are often highly degenerate (especially in large networks) due to the fact that their value is characterised completely by single-edge rates. The IAR-MDPAlg will ``push" the second path towards the optimal path (where the edges are shared) but must then move along alternative edges due to the edge-disjointedness restriction, i.e.~it is then ``pulled" away from the optimal path. These alternative edges may possess poor rates, but the algorithm will nonetheless use them because of its now unstructured method of path building, and the cost matrix does not contain enough information to discourage their use. The push-and-pull continues until $\bs{s}$ is reached, by which point the route has been compromised and possesses a poor bottleneck rate. This push-and-pull effect is only avoided when the optimal end-to-end route from $\bs{t}$ to $\bs{s}$ through a neighbouring node is edge-disjoint with the optimal route, which is rare in large-scale, highly connected networks.

As a result, the bottleneck rate cost functions used within DA are not so effective in the present context. %for the MDPAlg. 
One should avoid the employment of routing cost functions that possess high degeneracies, such as the direct translation of the widest path cost functions. Instead, one should explore cumulative costs, similar to that used in the shortest path formulation. Cumulative cost functions which capture properties of entire paths are more suited to this algorithm and can more effectively motivate end-to-end routing.

\subsection{Inverse-Accumulated-Rate Ansatz}

\begin{figure}[t!]
%\includegraphics[width=\linewidth]{end2end_waxman_thfib.pdf}
%Waxman Networks, $\sigma_r = 1$: \\ \hspace{1cm} \\ 
\includegraphics[width=\linewidth]{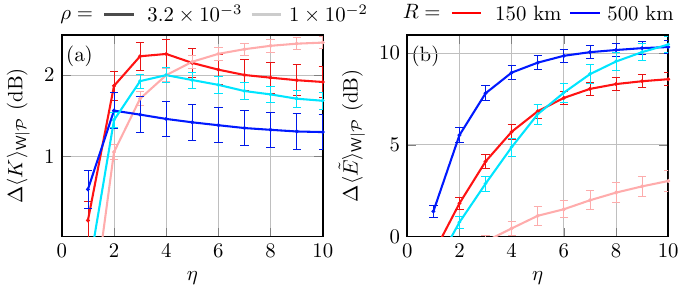}
%Scale-Free Networks, $\sigma_r = 1$: \\ \hspace{1cm} \\ 
\includegraphics[width=\linewidth]{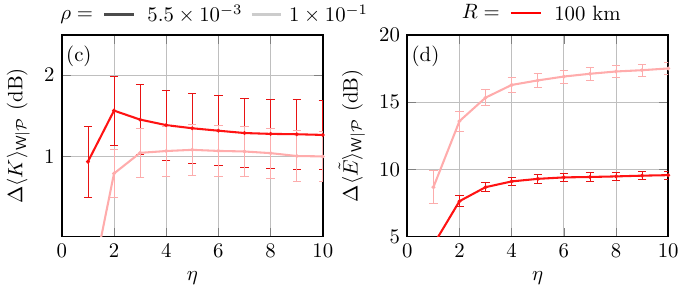}
\caption{Behaviour of IAR ansatz with respect to inverse rate penalty $\eta$ and fixed edge-usage penalty $\epsilon = 1$ over the classes of bosonic thermal-loss (a)-(b) Waxman networks and (c)-(d) scale-free networks ($\sigma_r=1$) using the thermal single-edge capacity upper-bounds. For a number of nodal densities $\rho$ and network radii, Panels (a),(c) plots the ensemble average rate amplification $\Delta \langle K \rangle_{\textsf{N}|\mc{P}}$ defined in Eq.~(\ref{eq:RateAdv}) gathered by the $M=2$ MDPAlg protocol using the inverse rate-sum ansatz, with respect to variable $\eta$. Panels (b),(d) plots the ensemble average consumption amplification $\Delta \langle \tilde{E} \rangle_{\textsf{N}|\mc{P}}$ defined in Eq.~(\ref{eq:ConsAdv}) with respect to $\eta$. 
}
\label{fig:IS_ansatz}
\end{figure}

%Fortunately, there are other heuristics that we may use to identify high-rate and resource efficient paths. The challenge with bottleneck rate maximisation and the MDPAlg emphasises the need for cumulative cost functions, those which measure accumulated cost along complete paths. Indeed, the shortest path algorithm is expediently employed in the MDPAlg framework for this reason. Keeping the task of rate-maximisation in mind we can be motivated by this and take an alternative approach. 

To achieve rate maximisation, a potential ansatz for the tentative cost function may take the form,
\begin{equation}
F_{\omega}^{\bs{s},{\bs{x}\rightarrow \bs{y}}}  =  T_{\bs{xx}} + K_{\bs{xy}}^{-\eta}+ \epsilon. \label{eq:IRS}
\end{equation}
which corresponds to minimizing the following global cost function,
\begin{align}
\mc{F}_{\omega} &= \sum_{(\bs{x},\bs{y}) \in \omega} (K_{\bs{xy}}^{-\eta} + \epsilon).
\end{align}
This aligns with minimizing the sum of the inverse point-to-point rates along a path $\omega$, adding a penalty for channel usage $\epsilon$ associated with each link. Minimizing this cost function simultaneously locates a path which maximizes the sum of the point-to-point rates along the path while minimizing its length; \textit{indirectly} identifying a high-rate and edge-efficient route.

In this tentative cost function, $\eta,\epsilon \in \mathbb{R}_{0}^+$ are hyperparameters used to impose a tradeoff between rate and path length. The parameter $\eta$ can be thought of as a rate motivator; if $\eta$ is large, then the inverse term $K_{\bs{xy}}^{-\eta}$ forces a large penalty to the cost function when low rate edges are included in the route. If the single edge rate is large, this incurs a low cost, motivating the use of an edge. Meanwhile, $\epsilon$ enforces a constant penalty term for edge usage, thus encouraging efficient routing. The edge-usage penalty can be generalised as an edge-wise property $\epsilon_{\bs{xy}}$ such as spatial length etc., to more accurately manage routing efficiency. The ability to control this tradeoff is extremely useful and is typically ignored in standard rate-maximisation algorithms.

With some simple numerical inspection, we see that the inverse rate-sum ansatz can achieve strong rate and resource management. In Fig.~\ref{fig:IS_ansatz} we perform an analysis of the ansatz given in Eq.~(\ref{eq:IRS}) for both Waxman networks and scale-free models. Here, we explore the minimum improvement offered via an MDPAlg protocol with two fixed routes, $\mc{P}_{\text{mdp}}^{M=2}$, by measuring the \textit{amplification} of its ensemble average rate and routing consumption over the optimal single-path protocol (measured in decibels),
\begin{align} 
\Delta \langle K \rangle_{\textsf{N}|\mc{P}} &\defeq 10 \log_{10} \left[ \langle K \rangle_{\textsf{N}|\mc{P}_{\text{mdp}}^{M=2}} / \langle K \rangle_{\textsf{N}|\mc{P}_{\text{sp}}}\right], \label{eq:RateAdv}\\
\Delta \langle \tilde{E} \rangle_{\textsf{N}|\mc{P}} &\defeq 10\log_{10} \left[ \langle \tilde{E} \rangle_{\textsf{N}|\mc{P}_{\text{mdp}}^{M=2}} / \langle \tilde{E} \rangle_{\textsf{N}|\mc{P}_{\text{sp}}}\right].  \label{eq:ConsAdv}
\end{align} 
When $\Delta \langle K \rangle_{\textsf{N}|\mc{P}}  > 0$ then the MDPAlg protocol offers a rate advantage over the optimal single-path protocol. However, the magnitude of $\Delta \langle \tilde{E} \rangle_{\textsf{N}|\mc{P}} > 0$ quantifies the \textit{increase} in routing consumption required to achieve this rate advantage. Optimally tuning the inverse rate-sum ansatz corresponds to maximizing $\Delta \langle K \rangle_{\textsf{N}|\mc{P}}$ while minimizing $\Delta \langle \tilde{E} \rangle_{\textsf{N}|\mc{P}}$.

Fig.~\ref{fig:IS_ansatz} illustrates the behaviour of these quantities with respect to a variable inverse-rate penalty $\eta$, while keeping $\epsilon = 1$ fixed and constant. For both classes of network, it is clear that for the scenarios considered in this work that the inverse rate-sum ansatz can be used to effectively enhance performance over that of single-path algorithms without incurring considerable routing cost. The algorithm can locate another end-to-end rate that effective enhances the rate while keeping the routing consumption increase at a manageable level. The more $\eta$ is increased, the greater emphasis it places on maximizing the end-to-end rate; however, this does not always correlate with \textit{obtaining} a greater end-to-end rate. Indeed, for both network models the rate enhancement tends to plateau while the routing consumption may continue to increase leading to wasted network usage. Hence, one should be careful not to increase $\eta$ too significantly with respect to the edge-usage penalty, $\epsilon$. 
%Furthermore, it is clear from our plots the increase of the rate enhancement offered by the $\mc{P}_{\text{mdp}}^{M=2}$ protocol diminishes with respect to 

Clearly, the optimal hyperparameters depend on the network model, connectivity and nodal density. An ideal implementation of the inverse rate-sum ansatz would fine-tune $\eta$ and $\epsilon$ corresponding to these properties. For simplicity,
throughout this work we have employed the quantities $\eta = 5$ and $\epsilon = 1$, motivated by the analyses above as a reasonable, multi-purpose choice with respect to different network characteristics. Future studies should focus on optimizing/controlling these hyperparameters with respect to specific network models or routing strategies. 

This is by no means an optimal approach. Nonetheless, the enormous benefit associated with quickly locating additional routes proves to outweigh any weakness in the approximation. 
In future investigations, it may be interesting to explore more sophisticated cost analyses, e.g.~a neural network variational ansatz. Such ansatzes may be of significant benefit when one wishes to optimise more than just rate and path length, e.g.~balancing the routing priorities of multiple user pairs, or considering waiting-times in quantum memories for entanglement distribution.


\begin{thebibliography}{44}
\bibitem{endnote_excess}
Note that, in continuous-variable (CV) QKD, thermal noise is used to model fiber imperfections that are not related to loss~\cite{Grangier03}. In fact, it is typical to introduce the channel's excess noise as $\varepsilon:=(1-\eta)\eta^{-1}\bar{n}_{\text{env}}$~\cite{TheorySim}. In experimental implementations, the typical value of the excess noise $\varepsilon$ is in the range $10^{-3}$--$10^{-1}$. For a transmissivity of $\eta=1/2$ (corresponding to $15$~km in standard fiber), we then have $\bar{n}_{\text{env}}$ in the same range of $10^{-3}$--$10^{-1}$.
\end{thebibliography}
\end{document}